\documentclass[acmtog]{acmart}

\usepackage[linesnumbered]{algorithm2e}
\usepackage{amsthm, amssymb, amsmath, latexsym, amsfonts, mathcomp, amscd}

\allowdisplaybreaks

\theoremstyle{plain}

\theoremstyle{remark}

\def\Z{\mathbb{Z}}
\def\R{\mathbb{R}}

\usepackage{mathtools}     
\usepackage{suffix}         
\usepackage{relsize}

\DeclarePairedDelimiterX\MeijerM[3]{\lparen}{\rparen}%
{\begin{smallmatrix}#1 \\ #2\end{smallmatrix}\delimsize\vert\,#3}     

\newcommand\MeijerG[8][]{%
  G^{\,#2,#3}_{#4,#5}\MeijerM[#1]{#6}{#7}{#8}}       

\WithSuffix\newcommand\MeijerG*[7]{%
  G^{\,#1,#2}_{#3,#4}\MeijerM*{#5}{#6}{#7}}

\def\BibTeX{{\rm B\kern-.05em{\sc i\kern-.025em b}\kern-.08emT\kern-.1667em\lower.7ex\hbox{E}\kern-.125emX}}

\copyrightyear{2019}
\acmYear{2019}
\setcopyright{acmlicensed}
\acmConference[Siggraph '19]{Siggraph '19}{}{}
\acmBooktitle{Siggraph '19}
\acmPrice{15.00}
\acmDOI{10.1145/1122445.1122456}
\acmISBN{978-1-4503-9999-9/18/06}

\begin{document}

%
\title{Local Fourier Slice Photography}

%
\author{Christian Lessig}
\authornotemark[1]
\email{lessig@isg.cs.uni-magdeburg.de}
\affiliation{%
  \institution{Institute for Simulation and Graphics, Otto-von-Guericke-Universit{\"a}t Magdeburg}
  \streetaddress{Universit{\"a}tsplatz 2}
  \city{Magdeburg}
  \postcode{39106}
  \country{Germany}
}

%
\renewcommand{\shortauthors}{Lessig}

%
\begin{abstract}
Light field cameras provide intriguing possibilities, such as post-capture refocus or the ability to synthesize images from novel viewpoints.
This comes, however, at the price of significant storage requirements.
Compression techniques can be used to reduce these but refocusing and reconstruction require so far again a dense pixel representation.
To avoid this, we introduce local Fourier slice photography that allows for refocused image reconstruction directly from a sparse wavelet representation of a light field, either to obtain an image or a compressed representation of it.
The result is made possible by wavelets that respect the ``slicing's'' intrinsic structure and enable us to derive exact reconstruction filters for the refocused image in closed form.
Image reconstruction then amounts to applying these filters to the light field's wavelet coefficients, and hence no reconstruction of a dense pixel representation is required.
We demonstrate that this substantially reduces storage requirements and also computation times.
We furthermore analyze the computational complexity of our algorithm and show that it scales linearly with the size of the reconstructed region and the non-negligible wavelet coefficients, i.e. with the visual complexity.
\end{abstract}

%
%


%

%
\begin{teaserfigure}
  \includegraphics[width=\textwidth]{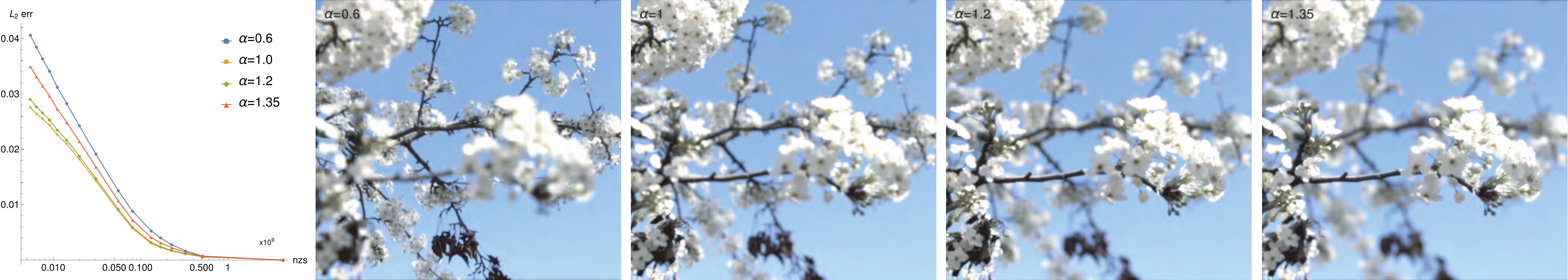}
  \caption{\emph{Right:} Refocused images obtained with local Fourier slice photography directly from a sparse wavelet representation of the light field.
    \emph{Left:} Reconstruction error as a function of the nonzero coefficients in the spare representation, demonstrating that the image fidelity degrades gracefully as storage requirements are reduced. The plot also shows the output sensitivity of our technique, with the largest error obtained when the in-focus region in the image is largest.}
  \Description{Light field photograph refocused with our technique. The graph also demonstrates the output-sensitivity of our technique with the largest error obtained when the in-focus areas are largest.}
  \label{fig:teaser}
\end{teaserfigure}

%
\maketitle

\section{Introduction}
\label{sec:introduction}

Light field cameras, which record the full four-dimensional plenoptic function, open up many possibilities for both consumer, e.g.~\cite{Ng2005a}, and professional applications, e.g.~\cite{Levoy2006a}.
Prime examples are post-capture refocus and the ability to obtain images where every depth is in focus.
The possibilities, however, come at the price of considerable storage requirements for the light field data sets.
Compression techniques can alleviate these but image reconstruction and light field processing typically require again a dense representation.

\begin{figure*}
  \includegraphics[width=\textwidth]{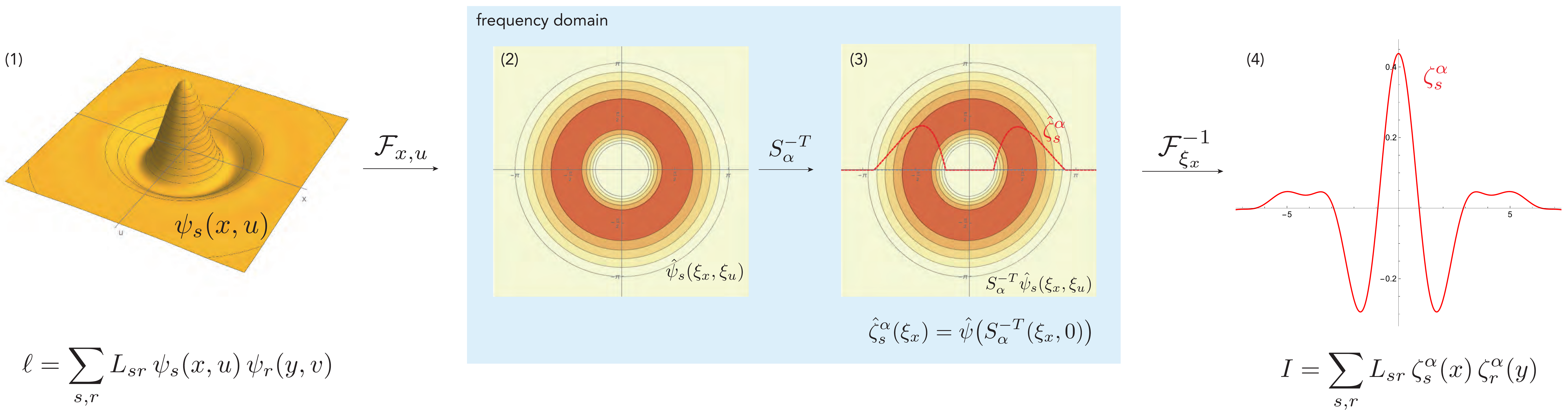}
  \caption{Overview of our approach: (1) A light field $\ell(x,y,u,v)$ in the two plane parameterization is represented using polar wavelets $\psi_s(x,u)$ and $\psi_r(y,v)$ (defined in Eq.~\ref{eq:polarlets:2d:hat}), exploiting the separability of the refocusing problem. (2) The wavelets are defined in polar coordinates in the Fourier domain. They are hence naturally compatible with the restriction to a line through the origin which implements refocused image reconstruction in the Fourier domain~\cite{Ng2005}. (3) The restriction of the sheared polar wavelet $\hat{\psi}_s( S^{-T} (\xi_x,\xi_u) )$ to the $\xi_x$ axis defines the one dimensional wavelet $\hat{\zeta}_{s}^{\alpha}(\xi_x) = \hat{\psi}_s( S^{-T} (\xi_x,0) )$ (and analogous for $(y,v)$). (4) The inverse Fourier transform of $\hat{\zeta}_{s}^{\alpha}(\xi_x)$, which can be computed in closed form, provides the exact, spatial reconstruction filters to obtain a refocused image from the wavelet coefficients $\ell_{sr}$ of the light field.}
  \label{fig:overview}
\end{figure*}

To avoid this, we propose \emph{local Fourier slice photography}, an algorithm to compute refocused images directly from a light field's sparse wavelet representation.
Our work draws inspiration from Ng's seminal Fourier slice photography~\shortcite{Ng2005} where image reconstruction is performed in the frequency domain using the projection slice theorem.
We combine this work with a recent advancement of the slice theorem~\cite{Lessig2018d} that uses carefully chosen wavelets to allow for an efficient projection from a signal's compressed wavelet representation. 
To apply this result to refocused image reconstruction, we extend it to a \emph{sheared, local projection slice equation} that establishes closed-form, shear-dependent reconstruction kernels for the projected signal.
With these, a refocused image can be obtained directly from a light field's compressed wavelet coefficients using an inverse transform.
We also derive an extension that enables one to directly obtain sparse, refocused images from a sparse light field data set.
Our experimental results confirm that our approach yields high fidelity, refocused images directly from compressed light fields without the need to obtain a dense pixel representation.
They also demonstrate that errors that arise at high compression rates mainly manifest themselves through lost high-frequency detail, i.e. without distracting artifacts.

The sparsity that reduces storage requirements also reduces computational costs.
We show this experimentally and verify it through a theoretical analysis that establishes a linear dependence on the number of nonzero wavelet coefficients.
Because of the spatial localization of the wavelets, the costs of our technique depend on the light field's angular resolution.
This was not the case for Fourier slice photography~\cite{Ng2005} although there one cannot easily take advantage of redundancy in the data.
The localization also enables us to obtain all-focus images, which is not possible using Fourier slice photography.

A conceptual overview of our approach is provided in Fig.~\ref{fig:overview} and the computations required in an implementation are summarized in Algorithm~\ref{algo:reconstruction}.
The remainder of the paper, which expounds on the details, is structured as follows.
After reviewing related work in the next section, we provide in Sec.~\ref{sec:polarlets} the necessary background on the polar wavelets that are used in our work.
In Sec.~\ref{sec:imaging} we derive the sheared local Fourier slice equation and develop our technique to obtain a refocused image directly from a light field's sparse wavelet representation.
Experimental results on refocused image reconstruction and all-focus images as well as details on our reference implementation are presented in Sec.~\ref{sec:experiments}.
We conclude in Sec.~\ref{sec:conclusion} with a discussion of possible directions for future work.

\section{Related Work}
\label{sec:introduction:related}

In computer graphics, light fields were introduced by Levoy and Hanrahan~\shortcite{Levoy1996} and Gortler et al.~\cite{Gortler1996}.
Initially, these were mainly of academic interest, e.g.~\cite{Chai2000}, but in the 2000s practical means to capture real-world light field data sets became available~\cite{Wilburn2005,Ng2005a,Venkataraman2013}.
With these, the generation, processing, and display of light fields has become an important research direction~\cite{Ihrke2016,Wu2017}.
In the following, we will therefore focus on related work most pertinent to our own.

Ng~\shortcite{Ng2005} showed that post-capture refocus can be formulated in the frequency domain using the Fourier projection slice theorem and that this provides an asymptotic speedup compared to the pixel domain.
Our work is inspired by Ng's and we combine it with a recent result in optics~\cite{Lessig2018d} that extends the  slice theorem to a spatially localized form using wavelets.
The extension relies on the use of polar wavelets, which is a family of wavelets defined separably in polar coordinates in the Fourier domain~\cite{Unser2010,Unser2011,Unser2013}.
These include a wide range of steerable wavelets~\cite{Perona1991,Freeman1991,Simoncelli1995} as well as curvelets~\cite{Candes2005a,Candes2005b} and ridgelets~\cite{Candes1999a,Donoho2000a}.
The present work also exploits the separability of polar wavelets in polar frequency coordinates and we extend~\cite{Lessig2018d} to include the shearing that implements refocusing.
We also benefit from the sparsity available with curvelet-like constructions~\cite{Candes2004,Donoho2000a}, which ensures an efficient sparse representation of light field data sets.

Light field imaging in the Fourier domain was also considered by Shi et al.~\shortcite{Shi2014}.
Their objective was to circumvent the sparsity degradation that results when the discrete instead of the continuous Fourier transform is used in numerical calculations.
This is no issue for our technique since our reconstruction kernels are obtained in the continuous Fourier domain.
Furthermore, while Shi et al. require a nonlinear optimization to obtain sparsity we use simple thresholding and rely on the compatibility of polar wavelets with the structure of natural images in frequency space~\cite[Ch. 9]{Candes2005a,Mallat2009}.
The design of of light field cameras and their sensors has been analyzed comprehensively by Liang and Ramamoorthi~\shortcite{Liang2015}.
Although these authors also perform their analysis in the Fourier domain, this work is orthogonal to ours since we assume we have a preprocessed light field data set as input.

For image synthesis, the frequency representation of the light field has been analyzed in a series of papers starting with the seminal work by Durand et al.~\shortcite{Durand2005}.
The work showed, for example, that the shearing that implements refocusing in the Fourier domain is the general expression for the transport of the light field in the two plane parametrization~\cite{Chai2000}.
Closest to our work are Fourier-based approaches for depth of field rendering~\cite{Soler2009a,Lehtinen2011a,Belcour2013}.
This work, however, aims at finding optimal sampling rates for the light field in Monte Carlo renderers while we assumes a (largely noise free) light field on the camera is available.
Because of the curse of dimensionality, image synthesis applications also typically do not employ an explicit basis representation of the light field, which is a key part of the present work.
Sen, Darabi, and Xiao~\cite{Sen2011a} used compressive sensing to reduce the sampling rate for depth of field rendering.
The polar wavelets employed in our work provide a sparse representation for image and light field data that would be well suited for compressive sensing.

Vagharshakyan, Bregovic, and Gotchev~\shortcite{Vagharshakyan2018} proposed the use of shearlets for light field reconstruction from a limited set of perspective views.
Shearlets can be seen as a stereographic projection of polar wavelets with the directional localization controlled by the parabolic scaling also used for curvelets.
The authors exploit the sparsity afforded by the shearlet transform to obtain an efficient algorithm for the reconstruction.
However, they do not exploit that image reconstruction is naturally formulated in polar coordinates in the frequency domain, which at the heart of our work.
In fact, to the best of our knowledge, with shearlets no closed form solution for the reconstruction kernels would be available.

Learning-based techniques for image reconstruction from light fields have also received considerable attention in recent work, e.g.~\cite{Levin2008,Kalantari2016,Yoon2015,Xu2018}.
The objective there is typically to perform the reconstruction from a reduced set of measurements, and it is hence orthogonal to our work.
In the spirit of~\cite{Vagharshakyan2018}, we believe that our polar wavelet representation of a light field might provide a useful pre-processing for learning-based reconstruction since it removes redundancy while respecting the intrinsic structure.

A variety of approaches for the compression of light field data sets have been proposed in the literature~\cite{Viola2017,Wu2017}, for example adapting techniques used for image compression, e.g.~\cite{Alves2018}; developing custom ones for slices or the full 4D light field, e.g.~\cite{Aggoun2006,Aggoun2008,Kundu2012,Conti2014}; or extending video compression schemes by exploiting that a light field can be seen as a sequence of images recorded from a set of nearby vantage points, e.g.~\cite{Vieira2015,Dai2015}.
Our use of polar wavelet for transform coding is dictated by our objective to refocus from the compressed representation.
However, we do not provide a full compressions technique, e.g. we do not consider the choice of color space, gamma correction, and quantization.

\section{Polar Wavelets}
\label{sec:polarlets}

\begin{figure}[t]
  \includegraphics[width=0.25\columnwidth]{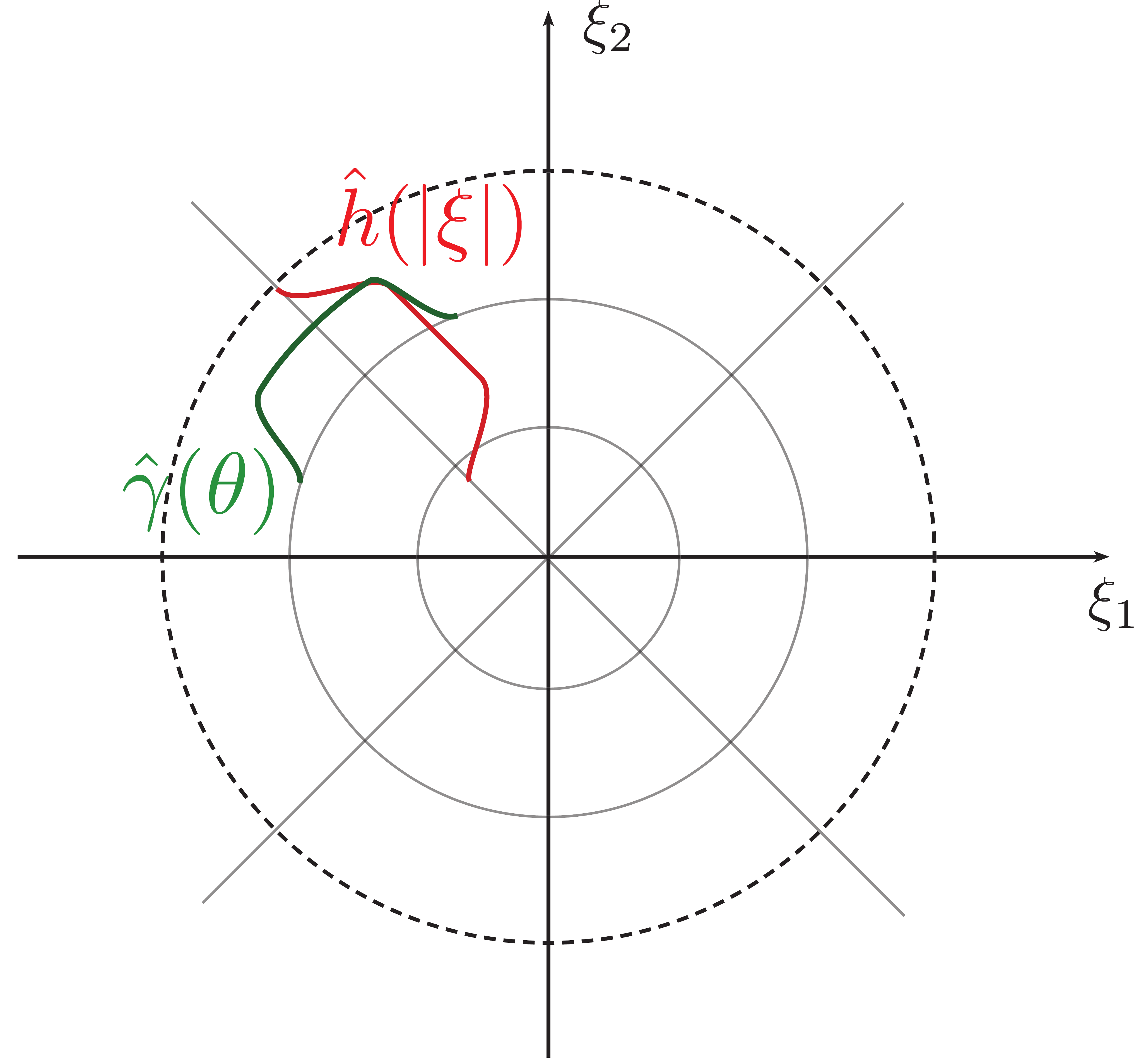}
  \includegraphics[width=0.35\columnwidth]{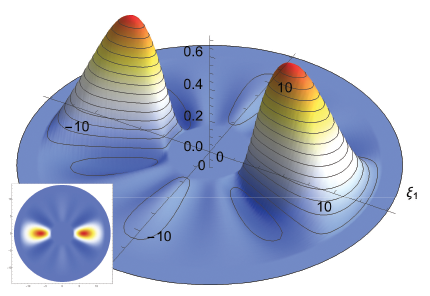}
  \includegraphics[width=0.35\columnwidth]{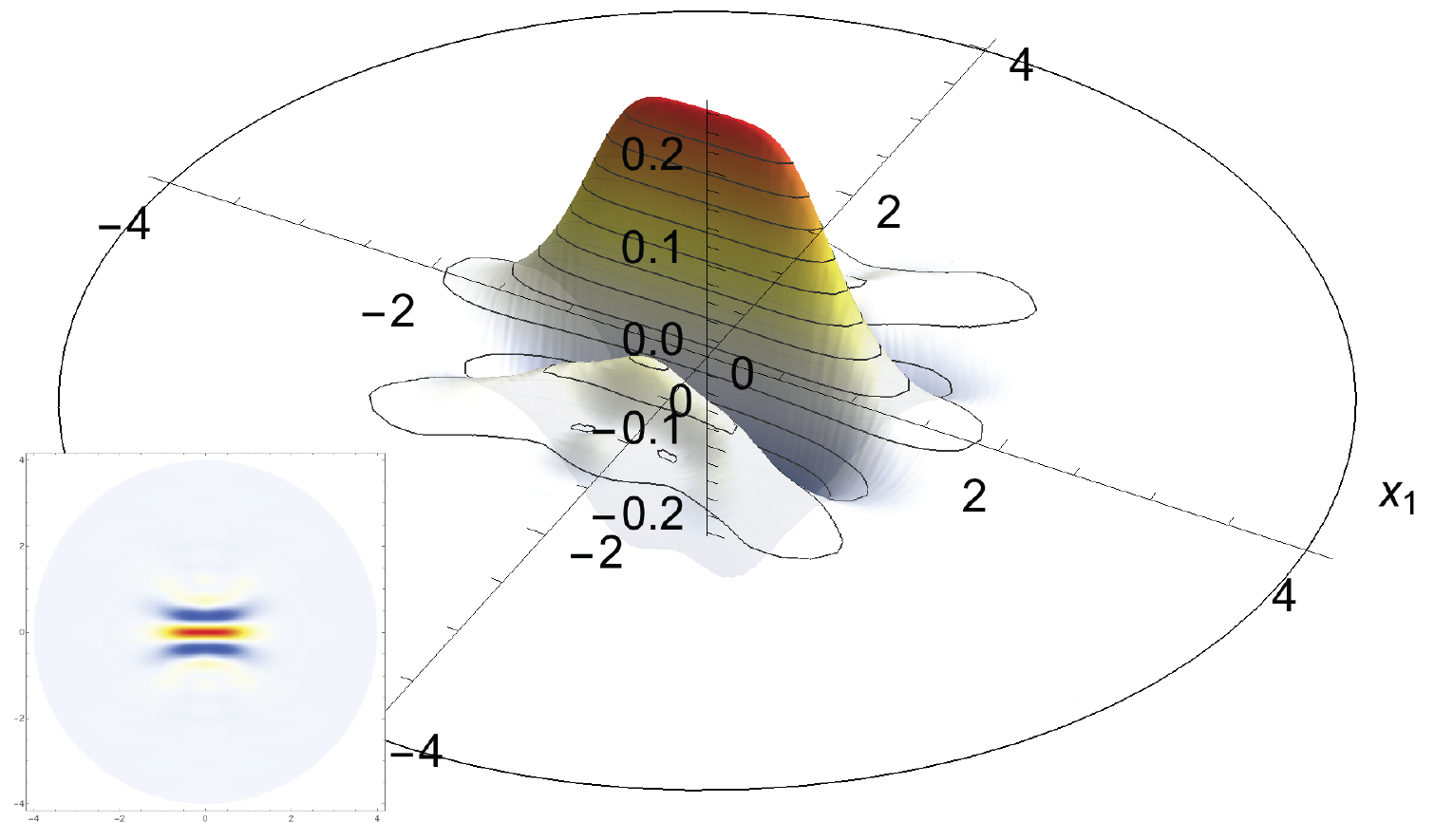}
  \caption{\emph{Left:} Conceptual construction of polar wavelets using window functions separable in polar coordinates. \emph{Right:} Directional, curvelet-like polar wavelets in the frequency domain (middle) and the spatial domain (right). }
  \label{fig:polarlets}
\end{figure}

Wavelets are functions that are well localized in both the spatial and frequency domain.
Through this, they enable an efficient and sparse representation of signals and can reduce the computational costs of numerical computations, e.g.~\cite{DeVore2006,Stevenson2009a}.
In multiple dimensions, wavelets are typically constructed as tensor products of one dimensional ones.
Polar wavelets~\cite{Unser2013}, in contrast, are defined separably in polar coordinates in Fourier space, which leads to many desirable properties.

Polar wavelets are constructed using a compactly supported radial window $\hat{h}( \vert {\xi} \vert)$, which controls the overall frequency localization, and an angular one, $\hat{\gamma}(\bar{\xi})$, which controls the directionality, with $\xi$ being the frequency variable and $\bar{\xi} = \xi / \vert \xi \vert$.
The mother wavelet is thus given by $\hat{\psi}({\xi}) = \hat{\gamma}(\bar{\xi}) \, \hat{h}( \vert {\xi} \vert )$, cf. Fig.~\ref{fig:polarlets}, left.
The whole family of functions used to represent arbitrary signals is then generated by dilation by $2^j$, $j \geq 0$, translation by $k \in \mathbb{Z}^2$, and rotation by $\theta_t$ with $t \in \{ 0, \cdots , T_j \}$.

The angular window $\hat{\gamma}(\bar{\xi})$ is conveniently described using a Fourier series in the polar angle $\theta_{\xi}$, i.e $\hat{\gamma}(\bar{\xi}) = \sum_{n} \beta_{j,n}^t \, e^{i n \theta_{{\xi}}}$.
In the frequency domain, a polar wavelet is thus given by
\begin{align}
  \label{eq:polarlets:2d:hat}
  \hat{\psi}_s({\xi})
  \equiv \hat{\psi}_{jkt}({\xi})
  = \frac{2^j}{2\pi} \Big( \sum_{n} \beta_{j,n}^t \, e^{i n \theta_{{\xi}}} \Big) \, \hat{h}(2^{-j} \vert {\xi} \vert ) \, e^{-i \langle {\xi} , 2^j {k}\rangle}
\end{align}
with the Fourier series coefficients $\beta_{j,n}^t$ controlling the angular localization.
In the simplest case, $\beta_n$ is the Kronecker delta $\delta_{n 0}$ and one has isotropic, bump-like wavelet functions, cf. Fig.~\ref{fig:overview}, left.
Conversely, when the support of the $\beta_{j,n}^t$ is over all integers $\Z$ then one can describe angular windows that are compactly supported in the polar angle $\theta_{\xi}$.
Eq.~\ref{eq:polarlets:2d:hat} then encompasses ridgelets~\cite{Candes1999a,Donoho2000a} and second generation curvelets~\cite{Candes2005b}, cf. Fig.~\ref{fig:polarlets}, right, which provide quasi optimally sparse representations for image-like signals.

\begin{figure}[t]
  \includegraphics[width=\columnwidth]{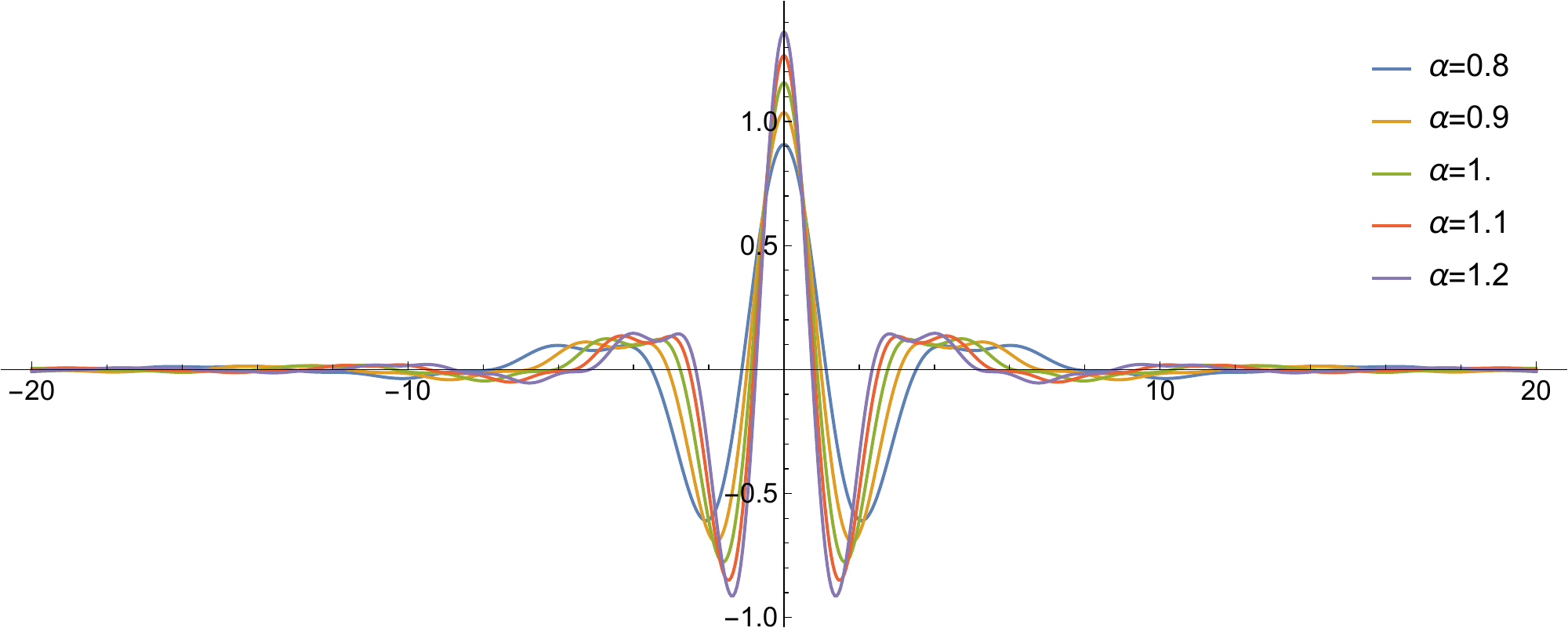}
  \caption{Reconstruction kernel $\zeta_{s}^{\alpha,0}(x)$ in Eq.~\ref{eq:slice_alpha:zeta_n:def} for different values of $\alpha$.}
  \label{fig:zeta_alphas}
\end{figure}

A beneficial property of polar wavelets is that their spatial representation, given by the inverse Fourier transform of Eq.~\ref{eq:polarlets:2d:hat}, can be computed in closed form.
Using the Fourier transform in polar coordinates one obtains~\cite{Lessig2018a}
\begin{align}
  \label{eq:polarlets:2d:space}
  \psi_{s}({x})
  \equiv \psi_{jkt}({x})
  = \frac{2^j}{2\pi} \sum_{n} i^n \, \beta_{j,n}^t \, e^{i n \theta_{{x}}} \, h_n( \vert 2^{j} {x} - {k}\vert )
\end{align}
where $h_n(\vert {x} \vert)$ is the Hankel transform of $\hat{h}(\vert {\xi} \vert)$ of order $n$.
For $\hat{h}(\vert {\xi} \vert)$ we employ the window proposed for the steerable pyramid~\cite{Freeman1991,Portilla2000} since $h_n(\vert {x} \vert)$ then has a closed form expression~\cite{Lessig2018a}.
Note that the angular localization of the wavelets, which is described by the $\smash{\beta_{j,n}^t}$ in Eq.~\ref{eq:polarlets:2d:hat} and Eq.~\ref{eq:polarlets:2d:space}, is invariant under the Fourier transform and only modified by the factor of $i^n = e^{i n \pi/2}$ that implements a rotation by $\pi / 2$.

As indicated in Fig.~\ref{fig:polarlets}, left, the radial mother window $\hat{h}(\vert \xi \vert)$ is defined away from the origin and dilation by $2^{-j}$ centers it at higher and higher frequencies as $j \geq 0$ grows.
To represent arbitrary signals, including those whose Fourier transform is nonzero around the origin, an additional window $\hat{g}(\vert \xi \vert)$ is required that has support in the disk-neighborhood around $\xi = 0$ (and complements $\hat{h}(\vert \xi \vert)$).
The translates of the inverse Fourier transform of $\hat{g}(\vert \xi \vert)$ yield the so called scaling functions $\phi_{k}({x})$ and in our case these will always be isotropic.
To simplify notation we will write $\psi_{-1,k}({x}) \equiv \phi_{k}({x})$; we refer to~\cite[Ch.~5]{Daubechies1992} for more details on the concept of scaling functions.

The polar wavelets in Eq.~\ref{eq:polarlets:2d:space} together with the just defined scaling functions provide a Parseval tight frame for $L_2(\mathbb{R}^2)$.
Thus any function $f({x}) \in L_2(\mathbb{R}^2)$ can be represented as~\cite{Unser2013}
\begin{align}
  \label{eq:polarlets:2d:frame}
  f({x})
  &= \sum_{j=-1}^{\infty} \sum_{k \in \mathbb{Z}^2} \sum_{t=1}^{T_j} \underbrace{\left\langle f({y}) \, , \, \psi_{jkt}({y}) \right\rangle}_{\displaystyle f_{jkt}} \, \psi_{jkt}({x}) .
\end{align}
where $\langle \, , \rangle$ is the $L_2$ inner product.
Although the above frame is redundant, since it is Parseval tight it still affords most of the conveniences of an orthonormal basis, e.g. the primary and dual frame functions coincide and the norm of the signal equals those of the expansion coefficients.
For an isotropic frame the redundancy is thereby $1+1/4+1/4^2+\cdots = 4/3$ and it increases when directional basis functions are used, i.e. with $T_j > 1$.
As for curvelets, where $T_j$ grows according to a parabolic scaling law as a function of $j$, anisotropic representations are typically sparser, which can compensate for the larger redundancy.

\begin{figure}[b]
  \includegraphics[width=\columnwidth]{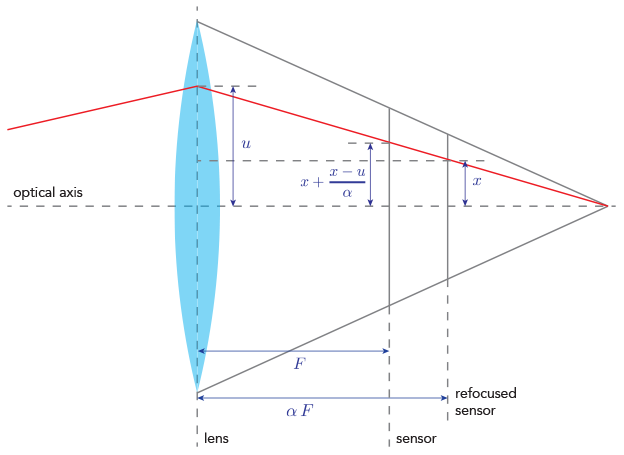}
  \caption{Image formation model used in our work. Shown is a cross section of the optical system with the red line representing a typical ray passing through the lens.}
  \label{fig:image_formation}
\end{figure}

Since the definition of polar wavelets in the frequency domain uses a compactly supported radial window $\hat{h}(\vert \xi \vert)$, the wavelets have non-compact support in space.
Thus, a finite signal representation is not exact, since it requires a truncation of the basis functions.
Nonetheless, with a sufficiently large apron region around an image, an arbitrary accuracy can be attained; in our experiments an apron of $4$ pixels sufficed to meet the requirements of photographic applications.
To obtain the wavelet representation of a signal, we compute it using a coarse-to-fine, fast wavelet transform-like algorithm where on each level the computations can either be performed using discrete convolutions with filter taps in the spatial domain or by multiplication with the windows in the frequency domain.

To simplify notation, we will typically employ the multi-index $s = (j,k,t)$ introduced in Eq.~\ref{eq:polarlets:2d:hat} and, when confusion might arise, write $s = (j_s,k_s,t_s)$. The index $s$ runs over the set $\mathcal{S}$ that a priori includes all scales, translations, and orientations. The cardinality of a set will be denoted by $\vert \cdot \vert$, e.g. $\vert \mathcal{S} \vert$.

\section{A Sheared Local Fourier Slice Equation for Computational Imaging}
\label{sec:imaging}

\begin{figure}
  \includegraphics[width=\columnwidth]{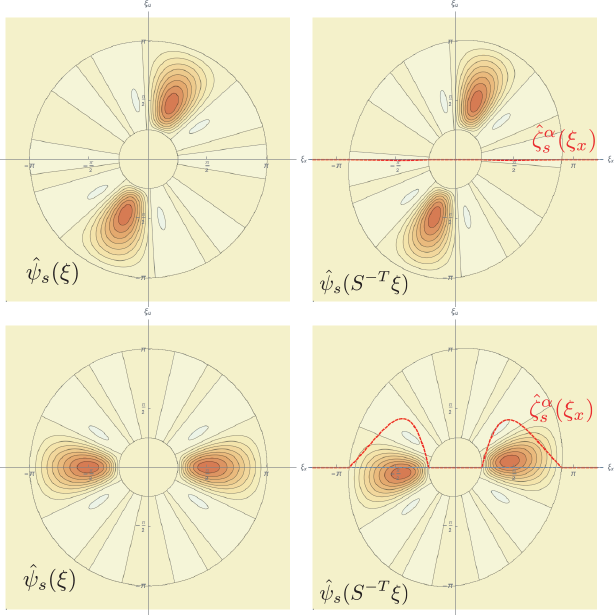}
  \caption{Contour plots of curvelet-like, directional polar wavelets in the frequency domain at two different orientations (different rows) and with and without shear (different columns) for $\alpha = 0.9$.
  Shown are also the sheared reconstruction filters $\hat{\zeta}_s^{\alpha}$.
  It is apparent that only those $\hat{\psi}_s(x)$ with an orientation close to the $\xi_x$-axis (dashed grey), i.e. to the slicing direction, yield $\hat{\zeta}_s^{\alpha}$ that contribute to the projected signal.
  }
  \label{fig:hw_aniso_sheared}
\end{figure}

In this section we derive local Fourier slice photography, our image reconstruction technique from a sparse polar wavelet representation of a light field.
We begin by fixing notation and recalling the image formation model.
At the end we analyze the computational complexity of the technique as well as sources of error.

\subsection{Image Reconstruction Model}
\label{sec:imaging:image_model}

We use the two-plane parametrization~\cite{Chai2000} for the light field $\ell(x,y,u,v)$ with $(x,y)$ being the coordinates on the image plane and $(u,v)$ those on the lens, see Fig.~\ref{fig:image_formation} for a schematic depiction.
We also assume $\ell(x,y,u,v)$ is non-zero only over the camera sensor in $(x,y)$ and over the lens in $(u,v)$ and that it already contains the $\cos{(\theta)}^4 / F^2$ foreshortening factor, where $\theta$ is the angle between the ray and the image plane normal.

For image formation we use the same model as~\cite{Ng2005}.
Hence, the image $I(x,y)$ is obtained from the light field as
\begin{align}
  I(x,y) = \frac{1}{\alpha^2} \! \int_{\R_{u}} \! \int_{\R_v} \! \! \ell \big(x/\alpha + (1 \! - \! 1/\alpha) u , y / \alpha + (1 \! - \! 1/\alpha) v, u, v \big) \, du \, dv
  \nonumber
\end{align}
where $\alpha = F / F'$ is the refocusing parameter, with $F$ and $F'$ being the original and new distance of the sensor to the lens plane, respectively, cf. Fig.~\ref{fig:image_formation}.
By changing the notation for the light field to $\ell( x,u ; y,v)$, the last equation can be written more compactly as
\begin{align}
  \label{eq:image_reconstruction:sheared}
  I(x,y) = \int_{\R_{u}} \! \int_{\R_v} \ell \big( S_{\alpha} (x,u)^T ; S_{\alpha} (y,v	)^T \big) \, du \, dv
\end{align}
where the shear matrix $S_{\alpha}$ is given by
\begin{align}
  \label{eq:image_reconstruction:shear}
  S_{\alpha} = \left(
  \begin{array}{cc}
    1/\alpha & 1-1/\alpha
    \\
    0 & 1
  \end{array}
  \right) .
\end{align}
The shear $S_{\alpha}$ amounts to the transport of the light field in the camera from the original sensor location to the refocused one~\cite{Chai2000,Durand2005}.

\begin{figure}
  \includegraphics[width=\columnwidth]{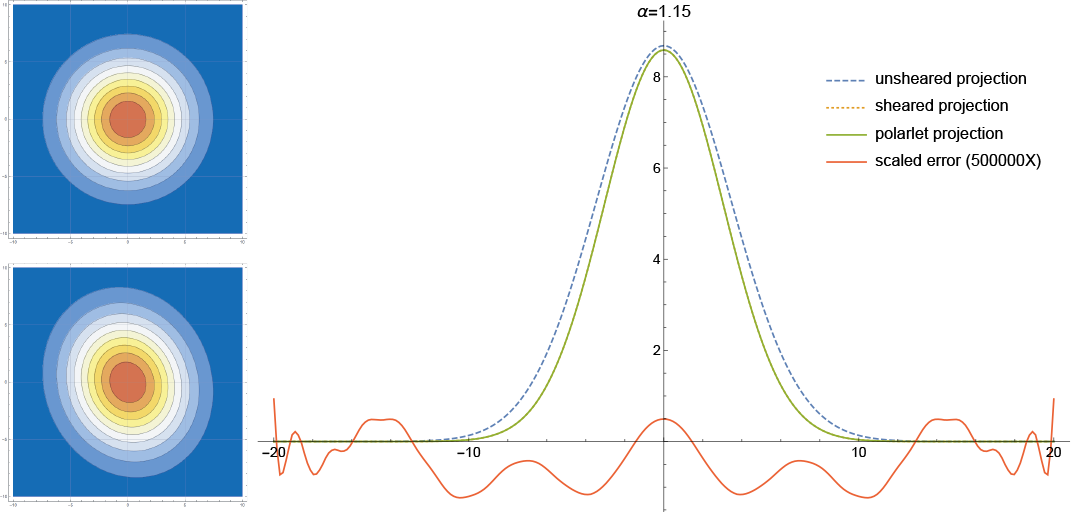}
  \caption{Projection of sheared 2D Gaussian (left, bottom). The maximum error with the sheared local Fourier slice equation is $1.86 \times 10^{-6}$.}
  \label{fig:gaussian_sheared_proj}
\end{figure}

To obtain a form of $\ell(x,u;y,v)$ that is amenable to compression, e.g. by thresholding small coefficients, we represent it in the polar wavelets introduced in Sec.~\ref{sec:polarlets}.
Respecting the separable structure of the refocusing in Eq.~\ref{eq:image_reconstruction:sheared}, that is performing one transform over $x$-$u$ and a second one over $y$-$v$, we obtain
\begin{align}
  \label{eq:lf:polarlets}
  \ell(x,u ; y,v) = \sum_{(s,r) \in \mathcal{L}} \ell_{sr} \, \psi_s(x,u) \, \psi_r(y,v)
\end{align}
where $\mathcal{L}$ is the index set for the representation that runs over all coefficients $s = (j_s,k_s,t_s)$ and $r = (j_r,k_r,t_r)$.
Inserting this representation into Eq.~\ref{eq:image_reconstruction:sheared} yields
\begin{align}
  \label{eq:lf:polarlets:2}
  I(x,y)
  &= \int_{\R_{u}} \! \int_{\R_{v}} \sum_{(s,r) \in \mathcal{L}} \ell_{sr} \psi_s\big( S_{\alpha} (x,u)^T \big) \, \psi_r \big( S_{\alpha} (y,v)^T \big)\, du \, dv
  \\[3pt]
  &= \sum_{(s,r) \in \mathcal{L}} \ell_{sr} \int_{\R_{u}} \! \psi_s\big( S_{\alpha} (x,u)^T \big) \, du \int_{\R_v} \! \psi_r \big( S_{\alpha} (y,v)^T \big) \, dv  .
  \nonumber
\end{align}
The last equation shows that it suffices to determine the effect of the sheared projection for the basis functions $\psi_s$ and $\psi_r$ and that this can be done independently for each of them.

Next, we will thus study general, two-dimensional sheared projection using polar wavelets.
This will yield our sheared local Fourier slice equation.
We return to imaging in Sec.~\ref{sec:imaging:slice_alpha:rendition}.

\subsection{A Sheared Local Fourier Slice Equation}
\label{sec:imaging:slice_alpha}

\begin{algorithm}[t]
\caption{Local Fourier slice photography algorithm for $\alpha$-sheared image reconstruction from wavelet compressed representation (for single color channel).}
\label{algo:reconstruction}
\SetKwProg{precomp}{Precomputation:}{}{end}
\SetKwProg{reconst}{Reconstruction:}{}{end}


\emph{// Input: Sampled light field $\ell$ in $(x,y) \times (u,v)$ parameterization}
\\
\precomp{($\ell$)}{

\BlankLine
\emph{// 1. Wavelet projection of light field in $(x,u)$ and $(y,v)$}
\\
$\ell_{\psi} = \{ \ell_{sr} \} = \mathrm{FWT}^2( \ell ) \in \R^{\vert S \vert \times \vert S \vert}$

\BlankLine
\emph{// 2. Computation of reconstruction filter, possibly sampling}
\\
\emph{// it for interpolation}
\\
$\zeta_s^{\alpha}(x) = \mathcal{F}_x^{-1} \left( \hat{\psi}_s\big( S_{\alpha}^{-T} (\xi_x,\xi_u)^T \big\vert_{\xi_u = 0} \big) \right)$
\BlankLine
}
\BlankLine

\emph{// Input: shear $\alpha$, resolution $N$ for reconstruction}
\\
\reconst{($\alpha$, $N$)}{

\BlankLine
\emph{// 1. Determine locations for reconstruction}
\\
$x = \big\{ -2^{-j_{\mathrm{max}}-1} N / \alpha, \cdots , 2^{-j_{\mathrm{max}}-1} N / \alpha \big\}$
\BlankLine

\emph{// 2. Evaluate projection of sheared locations}
\\
$k_s^{\alpha} = 2^{j_s} P_x(S^{-1} k_s)$
\BlankLine

\emph{// 3. Evaluate $\zeta_s^{\alpha}(x - k_s^{\alpha})$ for all $x_i$ and $k_s^{\alpha}$}
\\
$Z = \{ \zeta_s^{\alpha}(x_i - k_s^{\alpha}) \}_{i,s} \in \R^{n \times \vert \mathcal{S} \vert}$
\BlankLine

\emph{// 4. Reconstruction of $n \times n$ raw image}
\\
$I = Z \, \ell_{\psi} \, Z^T$
\BlankLine
}
\BlankLine
\end{algorithm}

Let $f(x,u)$ be a two dimensional signal.
We consider the sheared projection
\begin{align}
  g(x) = \int_{\R_u} f \big(S_{\alpha} (x,u)^T \big) \, du
\end{align}
where $S_{\alpha}$ is given by Eq.~\ref{eq:image_reconstruction:shear}.
When $f(x,u)$ is represented in polar wavelets the equation becomes
\begin{align}
  \label{sec:imaging:slice_alpha:2}
  g(x) = \sum_{s \in \mathcal{S}} f_s \, \int_{\R_u} \psi_s\big(S_{\alpha} (x,u)^T \big) \, du .
\end{align}
Inspired by Ng's work~\shortcite{Ng2005}, we will seek a numerically practical solution to Eq.~\ref{sec:imaging:slice_alpha:2} in the Fourier domain; a depiction of our approach is shown in Fig.~\ref{fig:overview}.
By the Fourier slice theorem, the integral in the last equation can be written as
\begin{align}
  \int_{\R_{u}} \! \psi_s\big( S_{\alpha} (x,u)^T \big) \, du
  =
  \mathcal{F}_x^{-1} \left( \alpha^{-1} \, \hat{\psi}_s\big( S_{\alpha}^{-T} (\xi_x,\xi_u)^T \big\vert_{\xi_u = 0} \big) \right)
\end{align}
where $S_{\alpha}^{-T} (\xi_x,\xi_u)^T \vert_{\xi_u = 0}$ is a linear slice in $\xi_x$-$\xi_u$ frequency space.
By expanding $\hat{\psi}_s$ using the definition in Eq.~\ref{eq:polarlets:2d:hat}, defining $\xi_0 = (\xi_x , 0 )^T$, and writing out the inverse Fourier transform we obtain
\begin{align}
  \label{eq:slice_alpha:zeta_n}
  \int_{\R_{u}} & \! \psi_s\big( S_{\alpha} (x,u)^T \big) \, du
  = \frac{2^j \, \alpha^{-1}}{(2\pi)^{3/2}} \sum_{n} \beta_{j,n}^t
  \\[4pt]
  &\times \underbrace{\int_{\R_{\xi_x}} e^{i n \theta_{S_{\alpha}^{-T} \xi_0}} \, \hat{h}\big( 2^{-j} \vert S_{\alpha}^{-T} {\xi}_0 \vert \big) \, e^{-i \langle {\xi_0} , 2^j S^{-1} {k}\rangle} \, e^{i \xi_x x } \, d\xi_x}_{\displaystyle \zeta_{s}^{\alpha,n}(x) = \zeta_{j_s,t_s}^{\alpha,n} \big( x - P_x(2^j S^{-1} {k_s}) \big)} ,
  \nonumber
\end{align}
that is, $\zeta_{n}^{s,\alpha}(x)$ is the inverse Fourier transform of the $\xi_x$-dependent term.
Importantly, for our choice of the radial window the $\zeta_{n}^{s,\alpha}(x)$ have a closed form expression, see Appendix~\ref{sec:appendix:zeta:spatial}, and the above equation can therefore easily be realized numerically.
Furthermore, the original shift $k_s$ of $\psi_s$ becomes after slicing $P_x(2^{-j_s} S^{-1} {k_s})$, where $P_x$ is the projection onto the $x$-axis.
This means the shape of the $\zeta_{n}^{s,\alpha}(x)$ remains independent of $k_s$.
Eq.~\ref{eq:slice_alpha:zeta_n} furthermore shows that the angular localization coefficients $\beta_{j,n}^t$ are invariant under the inverse Fourier transform.
The sheared projection of an arbitrary polar wavelet $\psi_x(x)$ is thus
\begin{align}
  \label{eq:slice_alpha:zeta_n:def}
  \zeta_{s}^{\alpha}(x) \equiv \frac{2^j \, \alpha^{-1}}{(2\pi)^{3/2}} \sum_{n} \beta_{j,n}^t \, \zeta_{s}^{\alpha,n}(x) = \int_{\R_{u}} & \! \psi_s\big( S_{\alpha} (x,u)^T \big) \, du .
\end{align}
and inserting Eq.~\ref{eq:slice_alpha:zeta_n:def} into Eq.~\ref{sec:imaging:slice_alpha:2} we obtain
\begin{align}
  \label{eq:slice_alpha:sheared_slice_eq}
   g(x)
   = \sum_{s \in \mathcal{S}} f_s \, \int_{\R_u} \psi_s\big(S_{\alpha} (x,u)^T \big) \, du
   = \sum_{s \in \mathcal{S}} f_{s} \, \zeta_{s}^{\alpha}(x) .
\end{align}
Eq.~\ref{eq:slice_alpha:sheared_slice_eq} is our sheared local Fourier slice equation with the $\zeta_s^{\alpha}(x)$ being the reconstruction filters that implement projection directly from the wavelet representation.
A simple verification of Eq.~\ref{eq:slice_alpha:sheared_slice_eq} for a two dimensional Gaussian, for which the ground truth has a closed form solution, is provided in Fig.~\ref{fig:gaussian_sheared_proj}.

\begin{figure}
  \includegraphics[width=\columnwidth]{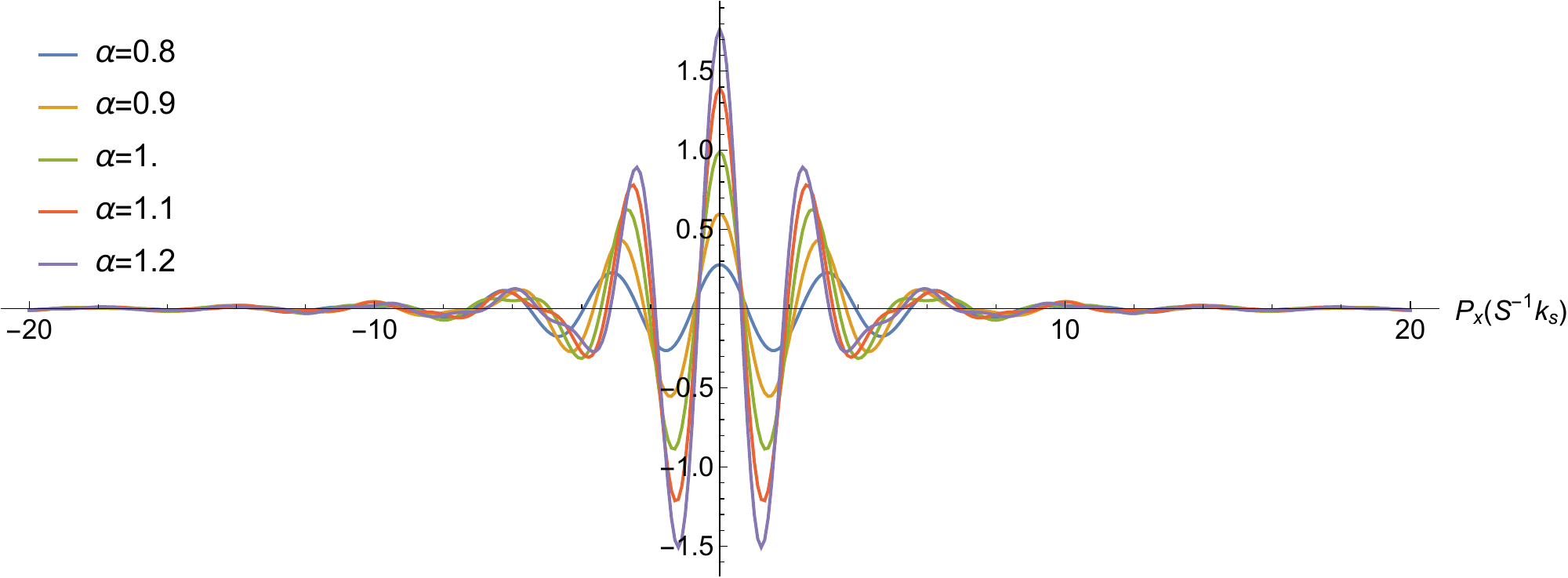}
  \caption{Coupling coefficient $\gamma_{sq}$ in Eq.~\ref{eq:compressed:gamma} for $j_s = j_q$ and $k_q = 0$ as a function of $P_x(2^{-j} S^{-1} k_s)$.}
  \label{fig:coupling_compressed}
\end{figure}

\begin{figure}
  \includegraphics[width=\columnwidth]{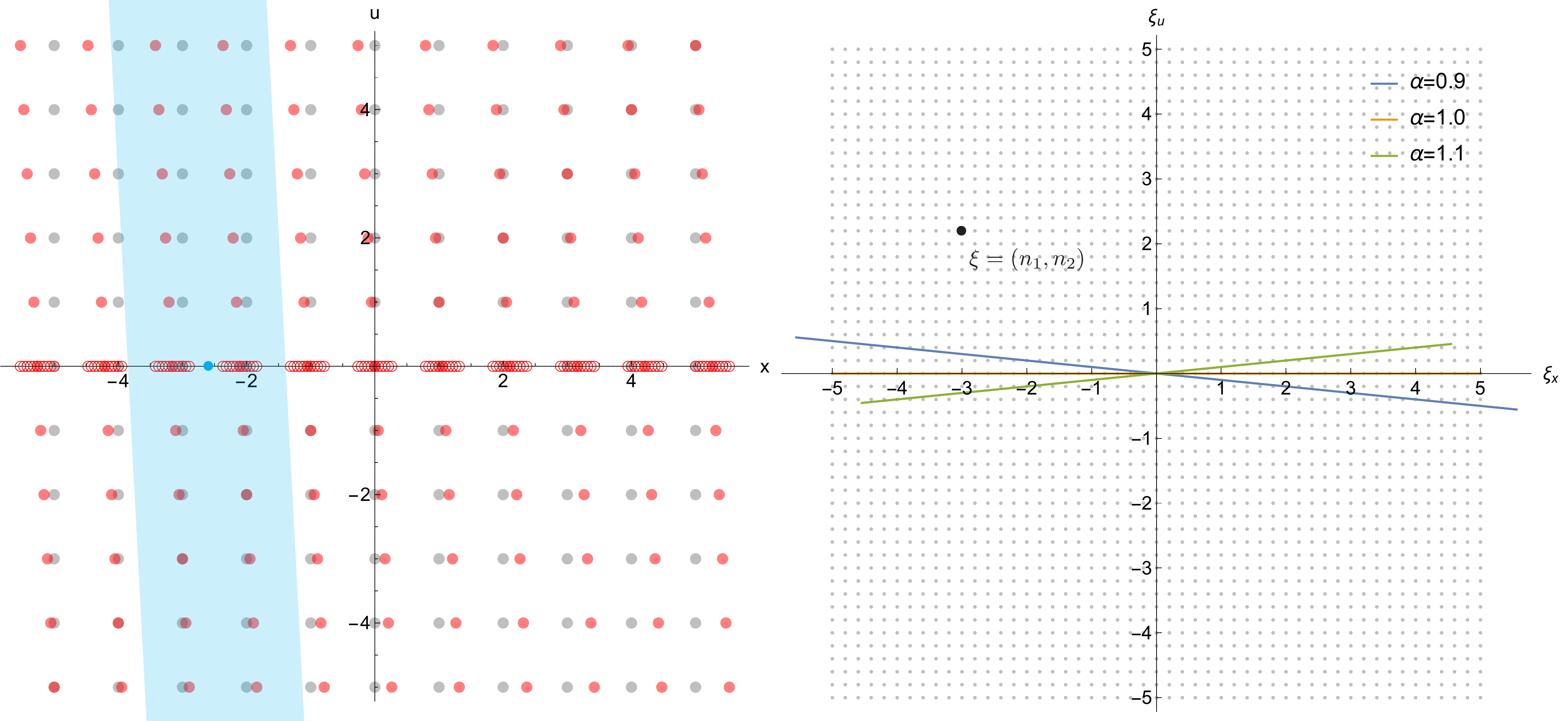}
  \caption{\emph{Left:} Original lattice of frame function locations (grey) and its shear (light red) for $\alpha = 0.95$. The projection (red circles) of the sheared lattice are on a denser grid, ensuring that the reconstruction filters $\zeta_s^{\alpha}$ are on a grid that is sufficiently fine to meet the Nyquist criterion for the increased bandlimit that arises through the shearing. The bluish region schematically indicates the set of coefficients that determines the projection at the bluish location on the $x$-axis, indicating the locality of our approach. \emph{Right:} The discrete Fourier transform in Fourier Slice Photography~\cite{Ng2005} yields a signal on a discrete lattice of frequencies (grey). The evaluation directions $S^{-T} \xi_0$ does not lie on the lattice and the resampling introduces error.}
  \label{fig:grid_sheared}
\end{figure}

The reconstruction kernels $\zeta_s^{\alpha}(x)$ are wavelet-like in that they are compactly supported in the frequency domain around $\xi_x = 2^{j_s}$ and well localized in the spatial domain.
The former holds since $\hat{\zeta}_s^{\alpha}(\xi)$ is a slice of a compactly supported wavelet centered at this frequency, cf. Fig.~\ref{fig:hw_aniso_sheared}, and the latter since the sliced window has the same decay as $\hat{\psi}_s(\xi)$, see Fig.~\ref{fig:zeta_alphas}.
The wavelet-like properties enable a local, sparse reconstruction with a coefficient $f_{s}$ only having a non-negligible effect to $g(x)$ in a small neighborhood around the projected locations $2^{-j_s} P_x(S^{-1} k_s)$.
Thus, only the wavelet coefficients defined at locations in a sheared tube in the $u$-direction above a location $x'$ contribute to $g(x')$, see Fig.~\ref{fig:grid_sheared} left.
We can therefore think of Eq.~\ref{eq:slice_alpha:sheared_slice_eq} as a wavelet representation of the projected signal with the wavelets $\zeta_s^{\alpha}(x)$ which are located at non-canonical locations $P_x(2^{-j_s} S^{-1} k_s)$.
In fact, with isotropic wavelets and $\alpha=1$ the projection in Eq.~\ref{eq:slice_alpha:sheared_slice_eq} becomes
\begin{subequations}
\begin{align}
   g(x)
   &= \sum_{j,k_{x}} \underbrace{\Big( \sum_{k_{u}} f_{j,k_x,k_u} \Big)}_{\displaystyle f_{j,k_x}} \, \zeta_{j,k_{x}}^{1}(x)  ,
  \end{align}
\end{subequations}
which is a standard, one dimensional wavelet representation of $g(x)$.
Our result then also coincides with those of~\cite{Lessig2018d}.

Next to the spatial position of $\psi_s(x)$, the contribution of a coefficient $f_{s}$ to the projected signal $g(x)$ also depends on the orientation of $\psi_s(x)$, or, equivalently, on the corresponding $\smash{\beta_{j,n}^t}$, cf. Eq.~\ref{eq:slice_alpha:zeta_n:def}.
As is apparent from Fig.~\ref{fig:hw_aniso_sheared}, the magnitude of the reconstruction filters $\zeta_s^{\alpha}(x)$ is non-negligible only when the direction $S_{\alpha}^{-T} \xi_0$ overlaps the effective support of the wavelets in the frequency domain.
When curvelet- or ridgelet-like wavelets are used, i.e when $\hat{\psi}_s(\xi)$ has strong directional localization, then only a sheared wedge or a small number of wedges from the polar tiling of the frequency plane have effective support over the direction.
Hence, only these directions need to be considered in the sum over $s$ in the sheared Fourier slice equation in Eq.~\ref{eq:slice_alpha:sheared_slice_eq}.

Our result relies on the use of polar wavelets that, through their definition in polar coordinates in frequency space, are compatible with the intrinsic structure of the projection, i.e. with a restriction to a line through the origin in the Fourier domain.
With tensor product wavelets, e.g. using Daubechies-type discrete wavelets, one would have a different, skew slicing through the axis-aligned frequency window for every $\alpha$.
The reconstruction kernel would then not have a closed form solution and, since the wavelets have no simple description, even determining them numerically would be difficult.


\begin{figure*}
  \includegraphics[width=\textwidth]{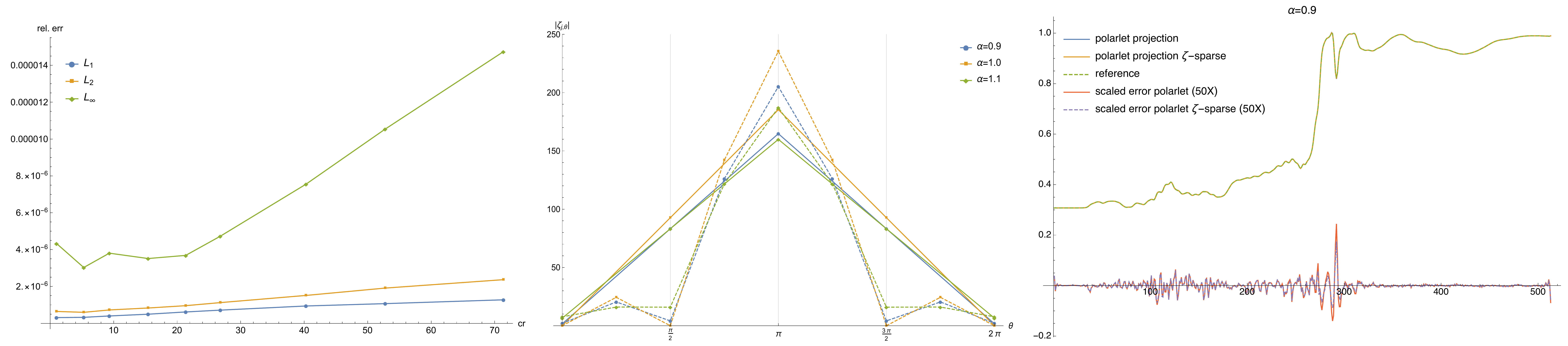}
  \caption{Sheared projection with anisotropic, curvelet-like wavelets. \emph{Left:} Reconstruction error as a function of the compression rate. \emph{Middle:} $\Vert \zeta_{jkt}(x) \Vert$ as a function of the orientation $t$; as expected from Fig.~\ref{fig:hw_aniso_sheared} the norm is non-negligible only when $t$ is approximately aligned with the projection direction. \emph{Right:} Reconstruction with all directions and only those where $\Vert \zeta_{jkt}(x) \Vert$ is non-negligible.}
  \label{fig:dragons_sheared_proj_aniso}
\end{figure*}

\subsection{Imaging with the Sheared Local Fourier Slice Equation}
\label{sec:imaging:slice_alpha:rendition}

We now return to refocused image reconstruction from the polar wavelet representation of a light field.
Using the local Fourier slice equation and inserting Eq.~\ref{eq:lf:polarlets:2} into Eq.~\ref{eq:slice_alpha:zeta_n:def} we obtain
\begin{align}
  \label{eq:slice_alpha:reconstruction}
   I(x,y)
   &= \sum_{(s,r) \in \mathcal{L}} \ell_{sr} \, \zeta_{s}^{\alpha}(x) \, \zeta_{r}^{\alpha}(y) .
\end{align}
The $\zeta_s^{\alpha}(x)$ thus provide the reconstruction filters that implement $\alpha$-refocused image reconstruction directly from the wavelet coefficients $\ell_{sr}$ of the light field, that is without the need to obtain a dense pixel representation.
In practice, a displayable representation of an image is obtained by evaluating Eq.~\ref{eq:slice_alpha:reconstruction} for every pixel (possibly with multiple samples to increase the quality).
We summarize the computations that are required for image reconstruction using Eq.~\ref{eq:slice_alpha:reconstruction} in Algorithm~\ref{algo:reconstruction}.

Eq.~\ref{eq:slice_alpha:reconstruction} can also be thought of as a separable wavelet representation of the image with the wavelets $\zeta_s^{\alpha}(x)$ and $\zeta_r^{\alpha}(y)$ located at the non-dyadic locations $P_x(2^{-j} S^{-1} k)$.
Note that the spatial and directional locality $\zeta_s^{\alpha}(x)$ discussed in Sec.~\ref{sec:imaging:slice_alpha} immediately carries over to Eq.~\ref{eq:slice_alpha:reconstruction} and, for example, a coefficient $\ell_{sr}$ contributes to the image $I(x,y)$ only when $\hat{\psi}_s(\xi_x,\xi_u)$ and $\hat{\psi}_r(\xi_y,\xi_v)$ are oriented along the slicing direction, cf. again Fig.~\ref{fig:hw_aniso_sheared}.

\subsection{Compressed Image Reconstruction}
\label{sec:imaging:slice_alpha_compressed}

It is often useful to directly determine the compressed representation of an image from a compressed light field, i.e. without first having to obtain a dense pixel representation as an intermediate step.
We will show next how the result of Sec.~\ref{sec:imaging:slice_alpha} can be extended towards this end.

We assume that the light field is again provided in the polar wavelet representation in Eq.~\ref{eq:lf:polarlets} so that the projected light field is given by Eq.~\ref{eq:slice_alpha:reconstruction}.
Assuming a separable wavelet basis $\psi_q^1(x) \, \psi_p^1(y)$ is used for the image, the corresponding expansion coefficients are given by
\begin{align}
  \ell_{qp} &= \Big\langle I(x,y) \, , \, \psi_{q}^1(x) \, \psi_p^1(y) \Big\rangle
  \\
  &= \sum_{(s,r) \in \mathcal{L}} \ell_{sr} \, \left\langle \zeta_{s}^{\alpha}(x) \, , \, \psi_{q}^1(x) \right\rangle_x \, \left\langle \zeta_{r}^{\alpha}(y) \, , \, \psi_{p}^1(y) \right\rangle_y .
  \nonumber
\end{align}
By introducing
\begin{align}
  \label{eq:compressed:gamma}
  \gamma_{sq} = \left\langle \zeta_{s}^{\alpha}(x) \, , \, \psi_{q}^1(x) \right\rangle
\end{align}
we can write this more compactly as
 \begin{align}
  \label{eq:compressed:final}
  \ell_{qp} &=  \sum_{(s,r) \in \mathcal{L}} \ell_{sr} \, \gamma_{sq} \, \gamma_{r p} .
\end{align}
When $\psi_{q}^1(x)$ and $\psi_{p}^1(y)$ are one dimensional, bandlimited ``polar'' wavelets that use the same radial window $\smash{\hat{h}(\xi_x)}$ as in Sec.~\ref{sec:polarlets}, then the $\gamma$-coefficients in Eq.~\ref{eq:compressed:gamma} can be computed in closed form; the expressions are provided in the accompanying Mathematica code.
By the compact support of the wavelets in the frequency domain, the $\gamma_{sq}$ are then, for moderate $\alpha$, non-negligible only when $\max( 0, \vert j_q - 1 \vert) \leq j_s \leq \vert j_q + 1 \vert$; when $\vert \alpha - 1 \vert$ is large then $\max( 0, \vert j_q - 2 \vert) \leq j_s \leq \vert j_q + 2 \vert$ holds.
A plot of the $\gamma_{sq}$ as a function of $P_x(2^{-j} S^{-1} k_s)$ for different values of $\alpha$ and $j_s = j_q$ is provided in Fig.~\ref{fig:coupling_compressed}.
As can be seen there, the coefficients have a shape similar to $\zeta_s^{\alpha}(x)$, since $\hat{\zeta}_s^{\alpha}(\xi)$
 is essentially a smoothed box function, and in particular they have the same spatial decay.

A consequence of Eq.~\ref{eq:compressed:final} is that sparsity in the wavelet representation of a reconstructed image is induced by sparsity in those of the light field.
In particular, for fixed $(q,p)$ only the $\ell_{sr}$ with $\vert k_q - P_x(S_{\alpha}^{-T} k_s) \vert \lesssim 2^{-j_q}$ and $\vert k_p - P_x(S_{\alpha}^{-T} k_r) \vert \lesssim 2^{-j_p}$ contribute to $\ell_{qp}$, since the wavelets decay in space with dyadic dilation from level to level.
Hence, when all $\ell_{sr}$ in the sheared tube above $k_q$ and $k_p$ are negligible, cf. Fig.~\ref{fig:grid_sheared}, then also $\ell_{qp}$ is negligible.
The coefficient $\ell_{qp}$ can also become small through cancellation, since both the $\ell_{sr}$ and $\gamma_{sq}$ are signed.
Intuitively, this happens, for example, when a region is defocused in the sheared projection and hence the wavelet coefficients on fine levels there have to be small.
Such decay estimates are beyond the scope of the present paper and will be investigated elsewhere; existing result in this direction can be found in~\cite{Quinto1993,Quinto2007}.

\subsection{Computational Complexity of Image Reconstruction}
\label{sec:imaging:complexity}

In the following, we will analyze the computational complexity of Algorithm~\ref{algo:reconstruction}.
We assume that the reconstruction is performed for a region $A \subseteq [0,1]^2$, which is potentially a subset of the normalized original image plane $[0,1]^2$, with size $\vert A \vert$ and that $R$ is the sampling rate per pixel in the reconstructed image.
The input light field is assumed to have resolution $N_x \times N_x \times N_u \times N_u \times 3$ (we will ignore the apron that is used in practice to avoid boundary artifacts, since it is small and has a negligible effect on the complexity yet would make the analysis considerably more cumbersome).

Step~1 and Step~2 in Algorithm~\ref{algo:reconstruction} have negligible costs and will be disregarded.
For Step~3, the separable reconstruction kernel $\zeta_s^{\alpha}(x) \, \zeta_r^{\alpha}(y)$ has effective support $2^{-j_s} W \times 2^{-j_r} W$, where $W$ is a window function-dependent constant (see Fig.~\ref{fig:zeta_alphas}).
The computational costs $K_{sr}$ that arise for each coefficient $\ell_{sr}$ are thus
\begin{align}
  \label{eq:math:complexity:zeta_eval}
  K_{sr} = 2 \, c_{\zeta} \, 2^{-j_s - j_r} W^2 \, R
\end{align}
where $c_{\zeta}$ is the cost for evaluating $\zeta_s^{\alpha}(x)$ at one point.
The number of coefficients on level $(j_s,j_r)$ in a dense wavelet representation is
\begin{align}
  \big\vert \mathcal{L}_{j_s,j_r} \big\vert = 4 \cdot \, 2^{-2(j_{max} - j_s)} \, 2^{-2(j_{max} - j_r)} \, N_x^2 \, N_u^2 \, T_{j_s} \, T_{j_r} \, \vert A \vert
\end{align}
since at each level the resolution is reduced by a factor of $2$ in each direction.
As before, $T_{j}$ is the number of different orientations on level $j$ and the factor of $4 = 3 \cdot 4/3$ accounts for the three color channels and the redundancy of the isotropic frame.
The computational costs $K_{j_s,j_r} = K_{sr} \,  \big\vert \mathcal{L}_{j_s,j_r} \big\vert$ for image reconstruction on level $(j_s,j_r)$ are for a dense polar wavelet representation thus
\begin{align}
  K_{j_s,j_r}
   &= 4 \cdot 2^{-4j_{max}} \, 2^{j_r + j_s} \, N_x^2 \, N_u^2 \, T_{j_s} \, T_{j_r} \, \vert A \vert \, c_{\zeta} W^2 \, R .
\end{align}
For the costs $K_{j_{max}}$ across all levels we then have
\begin{align}
  K_{{j_{max}}}
   &= 4 \cdot P_{j_{max}}  \, N_x^2 \, N_u^2 \, T_{j_s} \, T_{j_r} \, \vert A \vert \, c_{\zeta} W^2 \, R .
\end{align}
where $P_{j_{max}} = 2^{-4 {j_{max}}} (2^{{j_{max}}+1} \! - \!1 )^2$.

\begin{figure}
  \includegraphics[width=\columnwidth]{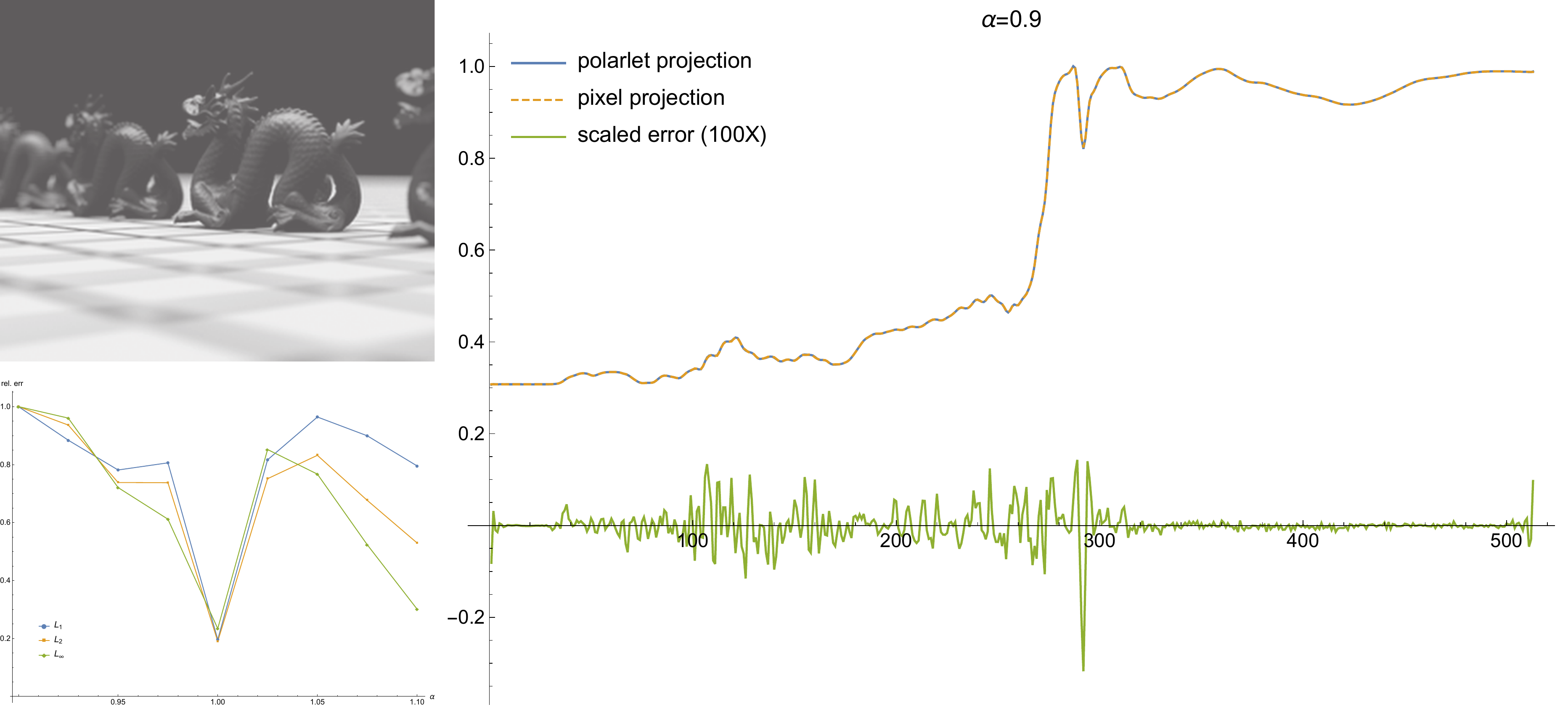}
  \caption{Sheared projection for the two dimensional dragon image in Fig.~\ref{fig:dragon_comparison_pixel} using classical and local Fourier slice photography.
  \emph{Right:} Projections and differences to pixel projection for a representative $\alpha$.
  \emph{Bottom left:} $L_1$, $L_2$, and $L_{\infty}$ errors of the projection as a function of the shearing angle $\alpha$.
  \emph{Top left:} Projection as a function of the interpolation order for classical Fourier slice photography.
  }
  \label{fig:dragons_sheared_proj}
\end{figure}

Two facts can be exploited to reduce the costs $K_{j_{max}}$.
First, with anisotropic, curvelet-like frame functions only the $\alpha$-sheared ones overlapping the slicing direction are required, cf. Fig.~\ref{fig:hw_aniso_sheared}.
These are of order one so that the costs $K_{{j_{max}}}^{\alpha}$ are
\begin{align}
  \label{eq:complexity:full}
  K_{{j_{max}}}^{\alpha}
   &= 4 \cdot c_T \, P_{j_{max}}  \, N_x^2 \, N_u^2 \, \vert A \vert \, c_{\zeta} W^2 \, R .
\end{align}
%
where $c_T = \mathcal{O}(1)$.
The increased redundancy of a directional representation system hence does not affect the costs since the number of frame functions overlapping the slicing direction remains constant.
The second reduction of the costs $K_{j_{max}}$ arises from the sparsity in a light field's wavelet representation, i.e. that only a sparse set $\smash{\mathcal{L}_{j_s,j_r}^{\alpha,\epsilon}}$, consisting of non-negligible ones with respect to a compression parameter $\epsilon$ (in the simplest case of hard thresholding, $\epsilon$ gives the threshold), suffices to represent the signal.
The ratio between the number of coefficients in the full and sparse representations is known as compression rate
\begin{align}
  \mathrm{cr}_{j_s,j_r}^{\epsilon} = \frac{\vert \mathcal{L}_{j_s,j_r} \vert}{\vert \mathcal{L}_{j_s,j_r}^{\alpha,\epsilon} \vert} .
\end{align}
Assuming it is independent of the level, i.e. $\mathrm{cr}_{j_s,j_r}^{\epsilon} =  \mathrm{cr}_{\epsilon}$, when sparsity and directionality are exploited the costs $K_{{j_{max}}}^{\alpha,\epsilon}$ then are
\begin{align}
  K_{{j_{max}}}^{\alpha,\epsilon}
   &= 4 \cdot c_T \, P_{j_{max}}  \, \frac{N_x^2 \, N_u^2}{cr^{\epsilon}} \, \vert A \vert \, c_{\zeta} W^2 \, R .
\end{align}
With basis dependent constants being ignored, this becomes in big-O notation
\begin{align}
  \label{eq:complexity:bigo}
  K_{{j_{max}}}^{\alpha,\epsilon} = \mathcal{O}\Big( \, P_{j_{max}}  \, \frac{N_x^2 \, N_u^2}{cr^{\epsilon}} \, \vert A \vert \, R \Big).
\end{align}
This shows that the costs scale linearly in the area that is to be reconstructed and the sampling rate and as $1/cr^{\epsilon}$ in the compression rate.
Thus, as the number of coefficients in the sparse representation $\mathcal{L}_{j_s,j_r}^{\alpha,\epsilon}$ decreases and the compression rate grows also the computational costs decrease.

The complexity of Ng's Fourier slice photography~\cite{Ng2005} is, in our notation, $\mathcal{O} \big( \vert A \vert \, R \, N_x^2 \big)$.
Although both works exploit that the projection becomes trivial in the Fourier domain, the spatial localization of the wavelets in our approach introduces again the directional resolution parameter $N_u$ in Eq.~\ref{eq:complexity:bigo}.
However, with the wavelets we also have a dependence on the compression rate $\mathrm{cr}^{\epsilon}$.
This can compensate for the $N_u$ factor, depending on the rate $\mathrm{cr}^{\epsilon}$ that can be attained for the light field.
Although theoretical characterizations of $\mathrm{cr}^{\epsilon}$ exist, see e.g.~\cite[Ch. 6, Ch. 9]{Mallat2009}, these require technical assumptions about the signal that are difficult to precisely meet in practice.
We will hence not pursue a further theoretical analysis here.
Nonetheless, the practical utility of wavelets for the compression of image like signals, and hence that significant compression rates can be attained, is by now well established, as is evidenced by their use in the JPEG2000 standard.

\begin{figure}[t]
  \includegraphics[width=\columnwidth]{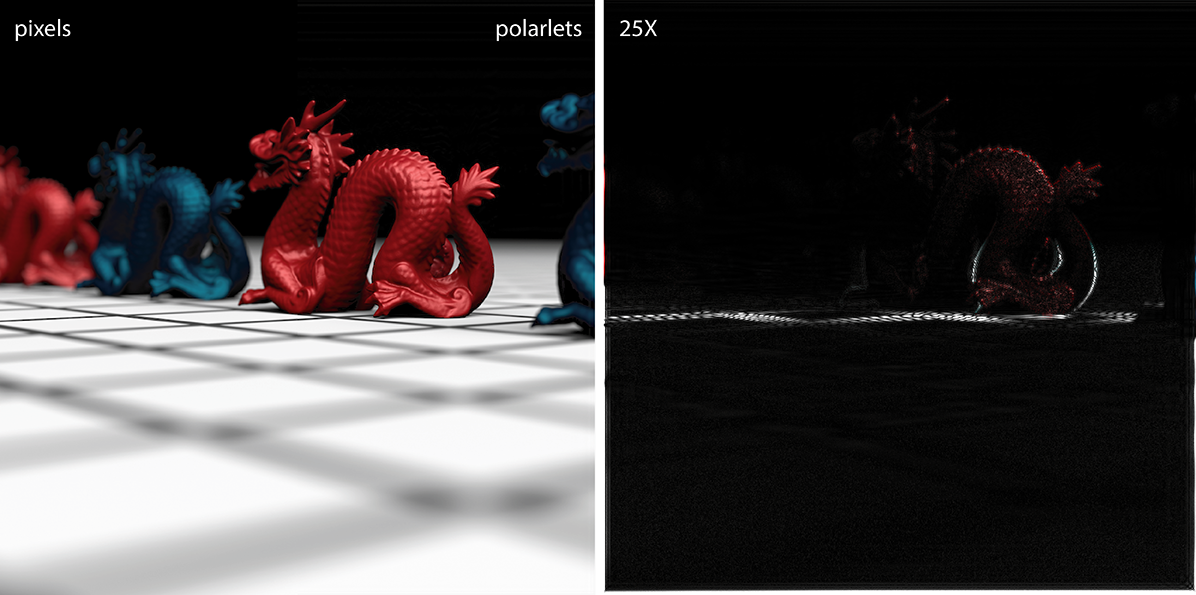}
  \caption{Comparison between our approach and reconstruction in the pixel domain for $\alpha = 0.9$. The left image is split in the middle between the pixel reconstruction (left) and the polar wavelet reconstruction (right). The image on the right shows the difference image, magnified by a factor of 25. The error is on the level of those incurred by na{\"i}ve slicing in the pixel domain.}
  \label{fig:dragon_comparison_pixel}
\end{figure}

\begin{figure*}
  \includegraphics[width=\textwidth]{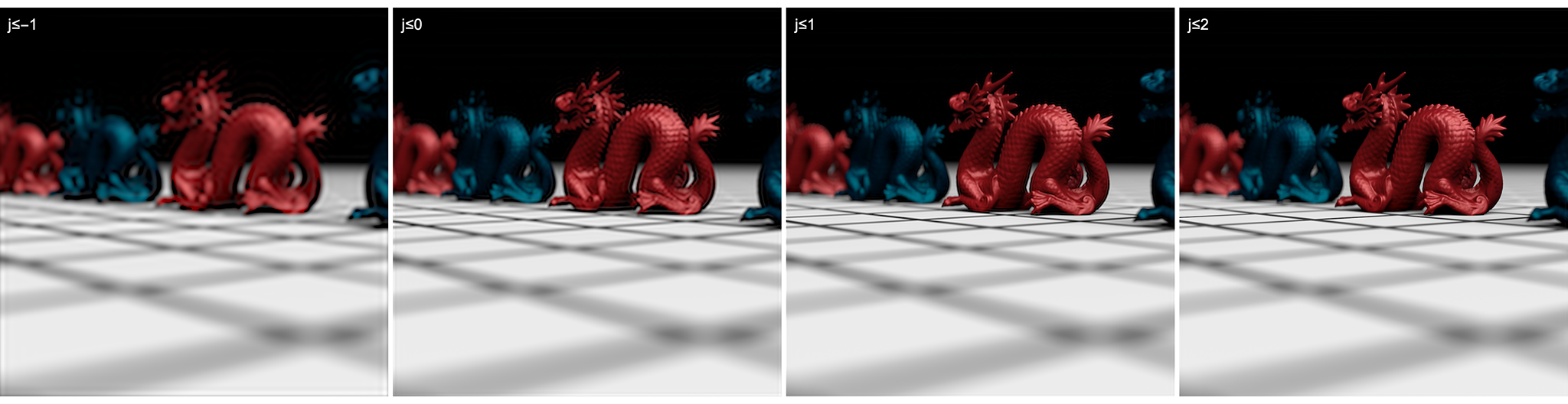}
  \caption{Reconstructions with subsets of all wavelet levels, demonstrating that not all are required to obtain acceptable reconstructions. Furthermore, the error increases gracefully as the number of levels decreases.
  The contribution by the individual levels is shown in Fig.~\ref{fig:dragons_rec_j_panel_44}.}
  \label{fig:dragons_rec_j_panel}
\end{figure*}

\subsection{Image Reconstruction Error}
\label{sec:math:error}

Ng~\shortcite{Ng2005} discusses two sources of error for image reconstruction, namely roll off error and aliasing.
We avoid roll-off error since our reconstruction kernels $\zeta_s^{\alpha}$ are the exact ones for our analysis wavelets, which in turn provide a Parseval tight frame; when the wavelets are used up to level $j_{\mathrm{max}}$ then all signals with bandlimit $2^j\pi$ can be represented exactly.
Aliasing is of concern because the shearing can increase the bandlimit (the change in the bandlimit can be seen, e.g. in bottom right plot in Fig.~\ref{fig:hw_aniso_sheared}).
In our approach, this implies that for $\alpha < 1$ the $\zeta_s^{\alpha}$ have a bandlimit beyond those of the original polar wavelets $\psi_s$.
Consequently, they also need to be defined over a finer grid than the original wavelets to allow for perfect reconstruction.
But, as shown in Fig.~\ref{fig:grid_sheared}, left, the shearing also affects the grid over which the basis functions are defined and through this the $\zeta_s^{\alpha}$ are inherently defined on a lattice that has the appropriate density.

Beyond roll off error and aliasing, a third source of error in fact arises in Ng's work~\shortcite{Ng2005}.
As depicted in Fig.~\ref{fig:grid_sheared}, right, the discrete Fourier transform yields a lattice of discrete frequencies in the Fourier domain.
Except in the trivial case when $\alpha = 1$, the slicing direction $S^{-T} \xi_0$ does not lie on the lattice and the resampling onto a regular grid along the slicing direction provides an additional source of error.
This error does not occur in our approach since our basis functions are defined in the continuous Fourier domain and we analytically compute the projection as a function of $\alpha \in \R$.

The main source of inaccuracies in our approach are in practice those introduced by a finite truncation of the basis functions or, equivalently, of the filter taps used in the fast transform.
As discussed in Sec.~\ref{sec:polarlets}, these can be ameliorated by using appropriate padding and filter sizes.

%
%

\section{Experiments}
\label{sec:experiments}

In this section we present experimental results that verify the practicality of the image reconstruction technique developed in the last section.
Additional results as well as the raw images for most of the presented figures are provided in the supplementary material.

\begin{figure*}
  \includegraphics[trim=0 75 0 75, clip, width=\textwidth]{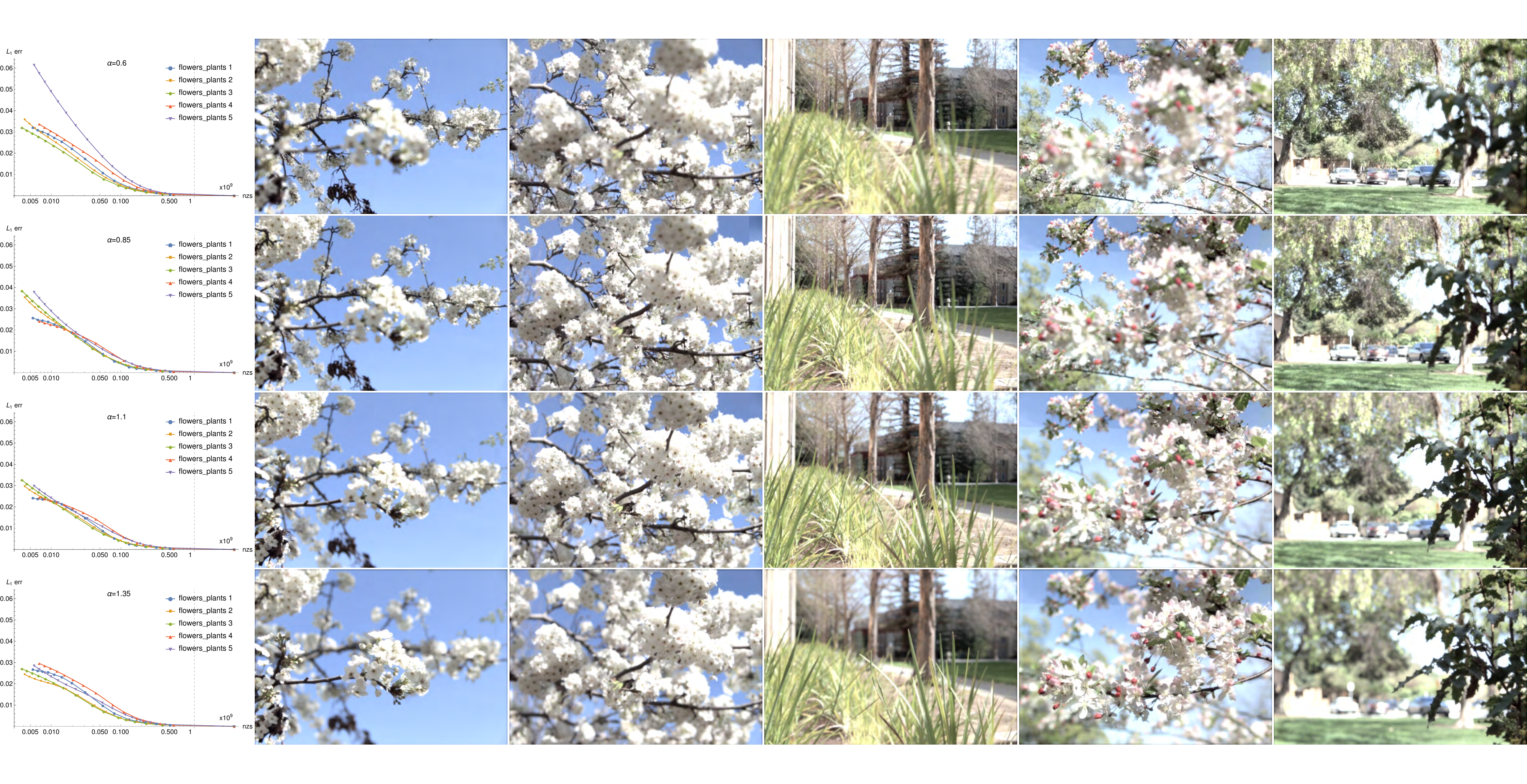}
  \caption{Log-linear plots of $L_1$ reconstruction error as a function of the nonzero coefficients for five different photographic light fields (flowers\_plants\_1 to flowers\_plants\_5 from left to right) and $\alpha = 0.6, 0.85, 1.1, 1.35$ (from top to bottom). The vertical grey dashed line in the plots on the left indicates the coefficients required for the uncompressed representation. Shown are reconstructed images obtained from the full wavelet representation to show the effect of changing $\alpha$ (see Fig.~\ref{fig:lf_th} for a depiction of compressed ones).}
  \label{fig:panel_general_comp_alpha_flowers_plants}
\end{figure*}

\subsection{Data sets}

We implemented a custom light field film class for the pbrt renderer~\cite{pbrt2} so that we could easily vary the spatial and angular resolutions as well as the optical properties of light field data sets.
The film class directly records light fields in the two plane parameterization, applies the foreshortening factor of $\cos{\theta}^4 / F^2$ and also computes a depth map for the scene.
We used pbrt to generated synthetic light fields for a dragon scene and a villa interior, both with a resolution of $1025 \times 1025 \times 33 \times 33 \times 3$.
We also performed experiments with photographic light fields from the Stanford Lytro light field archive.\footnote{\url{http://lightfields.stanford.edu/LF2016.html}}
We selected 20 light fields from different categories, all with a resolution of $541 \times 376 \times 14 \times 14 \times 3$.

All data sets were high dynamic range with computations performed in single precision.
We used two to three wavelet levels in all experiments (because of the limited resolution in the angular dimension) and for the experiments with the photographic light fields the same parameters (such as range of $\alpha$ values and tone mapping parameters) were applied in each case, which might be sub-optimal in individual instances.
Brightness variations that can be seen in some of the results for varying $\alpha$ stem from (independent) tone mapping.
%


\subsection{Implementation}

We developed a Mathematica reference implementation of Algorithm~\ref{algo:reconstruction} (available in the supplementary material) and a basic, multi-threaded C++ implementation.
Since the reconstruction filters $\zeta_s^{\alpha}(x)$ are relatively expensive to evaluate, cf. Appendix~\ref{sec:appendix:zeta:spatial}, we sampled them in a preprocessing step and interpolated at runtime (the error introduced through the interpolation was below $10^{-7}$ and thus negligible for photographic applications).
Post-processing bias was avoided by using one sample per pixel for image reconstruction and no interpolation filter on the image plane.
``Reference'' solutions were similarly computed using na{\"i}ve projection in the pixel domain without filtering of the light field data sets or the pixel data.

For the polar wavelets, we used filter taps of size $81 \times 81$ and, as mentioned earlier, an apron of 4 pixels. Larger values did not improve the reconstruction. Note that our algorithm for refocused image reconstruction is itself parameter free.

To study the effect of sparsification on image reconstruction quality we implemented a simple transform coding scheme with hard thresholding, i.e. we set to zero all coefficients whose magnitude is below a threshold $\epsilon$ dependent on the light field's $L_2$ norm to compensate for overall brightness differences between the data sets.
Results will be reported using either the number of nonzero coefficients in the light fields $\epsilon$-sparse wavelet representation (denoted by nzs) or the compression rate (denoted by cr), i.e. the number of nonzero coefficients in a sparse over the total number in a dense wavelet representation.
The number of nonzero coefficients provides an indication of the storage requirements although it is an upper bound since our wavelet representation only provides the transform coding step of a full compression scheme and we do not consider blocking, quantization, entropy coding and other aspects that are critical in practical compression algorithms.

\subsection{Simple experiments}

To demonstrate the correctness of the sheared local Fourier slice equation as well as to gain some understanding of various conceptual aspects we performed experiments on two dimensional signals yielding a one dimensional projection.

\paragraph{Basic verification}

We verified the correctness of the local Fourier slice equation using the sheared projection of a two dimensional Gaussian for which an analytic solution exists.
As shown in Fig~\ref{fig:gaussian_sheared_proj}, our reconstruction matches the analytic one very well with a maximum error on $1.86 \times 10^{-6}$.
This is of the same order as the reconstruction error of the 2D input signal, and hence attributable to inaccuracies in the transform yielding the wavelet coefficients.

\begin{figure*}
  \includegraphics[trim=0 75 0 75, clip, width=\textwidth]{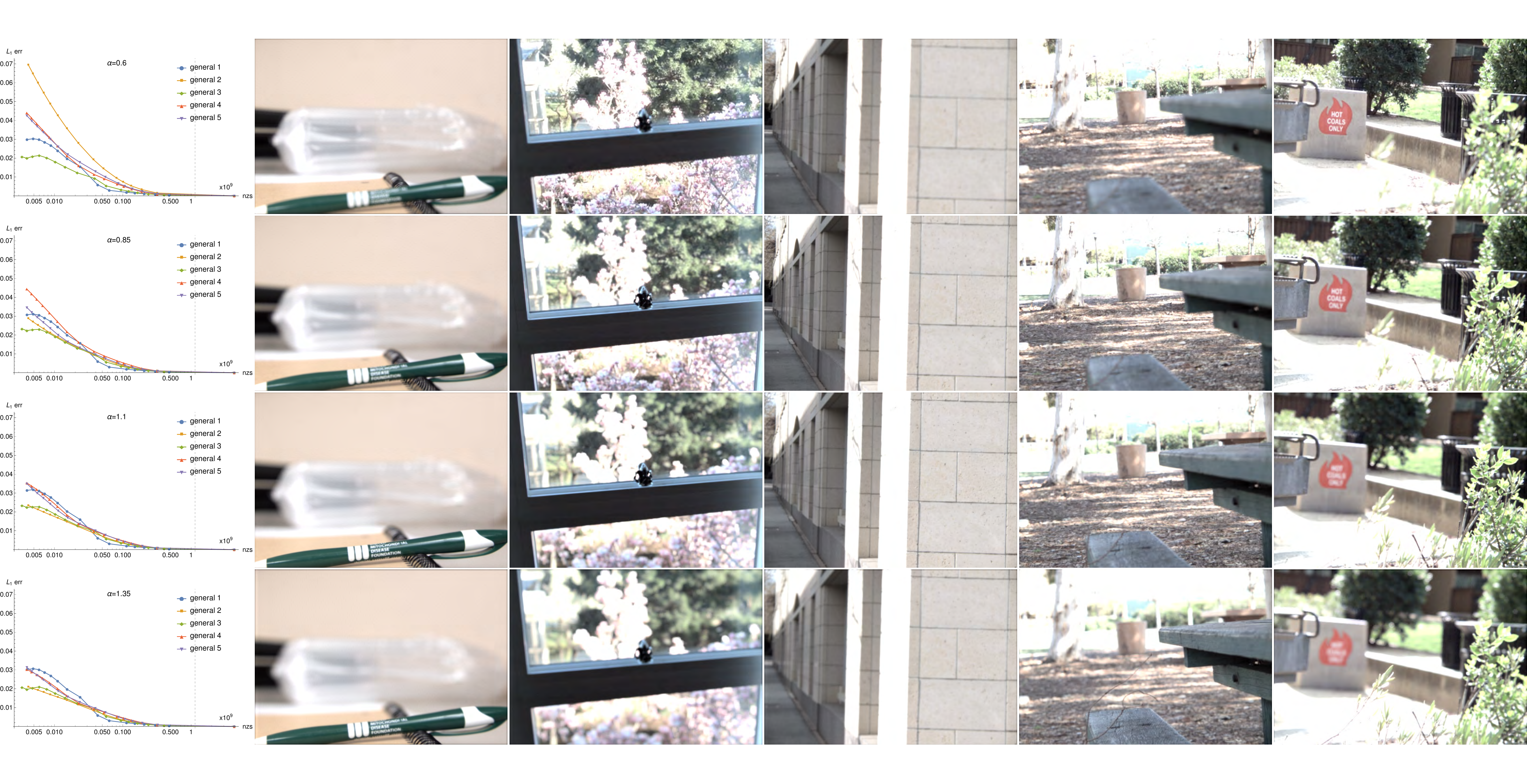}
  \caption{Log-linear plots of $L_1$ reconstruction error as a function of the nonzero coefficients for five different photographic light fields (general\_1 to general\_5 from left to right) and $\alpha = 0.6, 0.85, 1.1, 1.35$ (from top to bottom). The vertical grey dashed line in the plots on the left indicates the coefficients required for the uncompressed representation. Shown are reconstructed images obtained from the full wavelet representation to show the effect of changing $\alpha$ (see Fig.~\ref{fig:lf_th} for a depiction of compressed ones).}
  \label{fig:panel_general_comp_alpha_general}
\end{figure*}

\paragraph{Non-smooth signals and effect of $\alpha$}
To obtain insights on the behavior of our technique for ``natural images'' as well as to understand the effect of $\alpha$ on the reconstruction quality we considered the projection of a monochromatic dragons scene image (derived from the image in Fig.~\ref{fig:dragon_comparison_pixel}). 
The results in Fig.~\ref{fig:dragons_sheared_proj}, right, demonstrate that the error is smaller than what can be perceived visually and on the same order as differences resulting with different reconstruction kernels for the pixel domain projection (not shown).
Shown in the figure is also a quantitative analysis of the error as a function of $\alpha$, demonstrating that only a mild dependence on the angle exists.
The results, furthermore, reveal that our technique provides slightly lower error rates than classical Fourier slice photography~\cite{Ng2005}.
The plot on the top left of Fig.~\ref{fig:dragons_sheared_proj} indicates that the higher errors for the technique result from the interpolation from the axis-aligned DFT grid that is required for it, see Fig.~\ref{fig:grid_sheared}.
Since our wavelets are defined in the continuous Fourier domain and the restriction to a slice is performed there, no such interpolation is required.

\paragraph{Sparsity and Directionality}

For the dragon scene image we also studied the effect of sparsity as well as the gains that are possible using oriented, curvelet-like frame functions.
The left plot in Fig.~\ref{fig:dragons_sheared_proj_aniso} shows the error as a function of the compression rate with hard thresholding.
As one would expect for a wavelet representation, very accurate reconstructions are possible with a small fraction of the full coefficient set.
Furthermore, the error increases smoothly with the compression rate.
The middle plot in Fig.~\ref{fig:dragons_sheared_proj_aniso} depicts the norm $\Vert \zeta_{jkt}(x) \Vert$ of the reconstruction kernel $\zeta_{jkt}(x)$ as a function of the orientation $t$ of the wavelets.
Because of the localized support of the directional wavelets in the frequency domain, cf. Fig.~\ref{fig:hw_aniso_sheared}, only $\zeta_{jkt}(x)$ whose orientation $t$ matches the slicing direction $\smash{S_{\alpha}^{-T} \xi_0}$ are non-negligible.
The right plot in Fig.~\ref{fig:dragons_sheared_proj_aniso} verifies that orientations far from the projection direction do not have to be taken into account for reconstruction, i.e. the reconstruction error is sufficiently small when these are ignored.
Furthermore, since for typical values $\alpha$ does not fundamentally change the direction, some orientations are irrelevant, e.g. $t \leq \pi /2$ and $t \geq 3\pi /2$ in the plot.
Thus, the frame coefficients for the negligible orientations do not have to be stored, even if these are above an $\epsilon$-threshold.
$\Vert \zeta_{jkt}(x) \Vert$ can hence be understood as a signal independent form of sparsification.

\subsection{Image reconstruction}

\paragraph{Refocusing}

The reconstruction of a refocused image is provided in Fig.~\ref{fig:dragon_comparison_pixel}.
The comparison to the projection in the pixel domain, also shown in the image, verifies that our result is visually indistinguishable and that the differences between both are on the order of what one would obtain with different reconstruction kernels for projection in the pixel domain.

\begin{figure}[t]
  \includegraphics[width=\columnwidth]{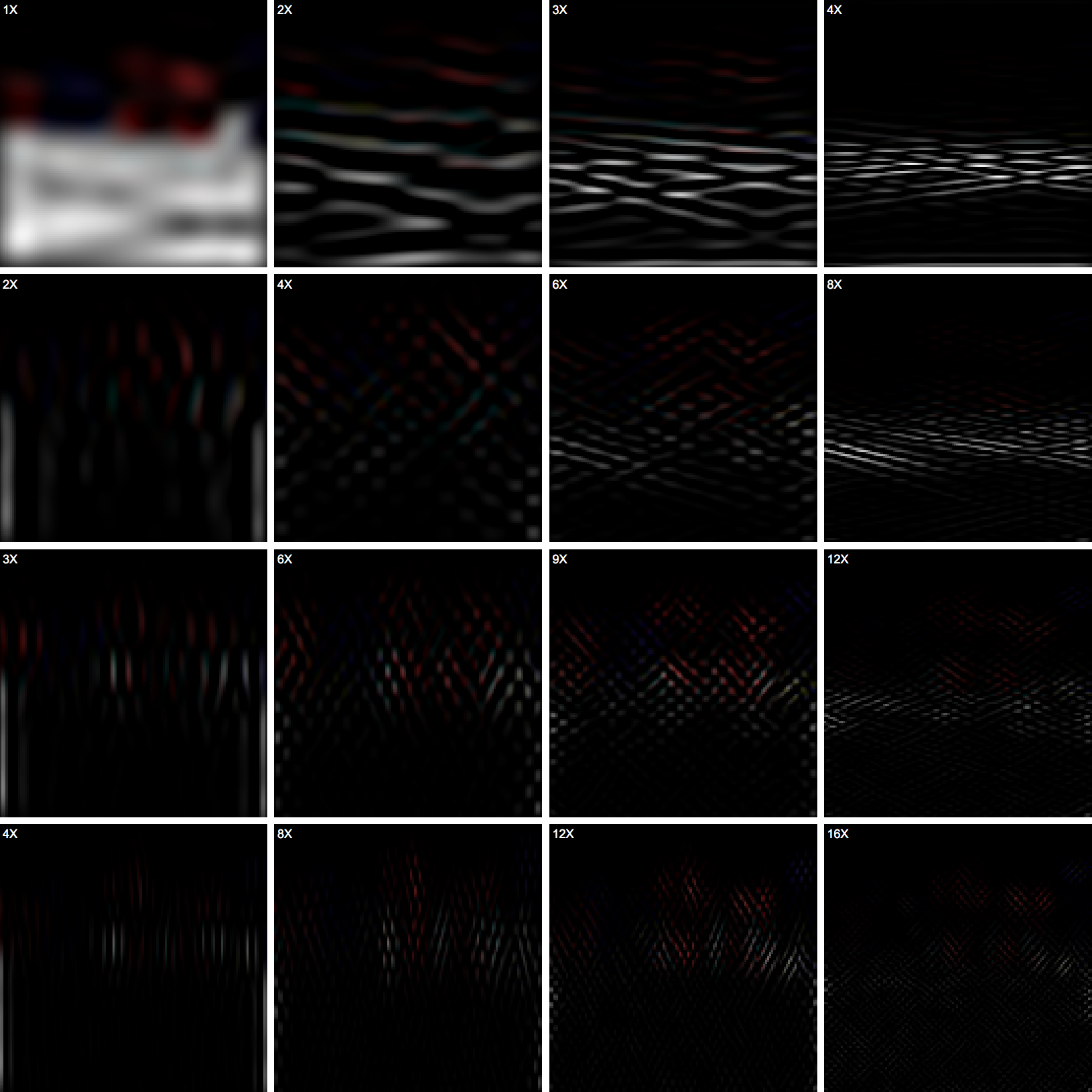}
  \caption{For the dragon scene in Fig.~\ref{fig:dragons_rec_j_panel}, contributions by different levels for $j_s = -1, \cdots, 2$ (columns) and $j_r = -1, \cdots, 2$ (rows) magnified by the factor shown in the inset. The results verify that very fine levels can be omitted when performance is of importance.}
  \label{fig:dragons_rec_j_panel_44}
\end{figure}

Fig.~\ref{fig:dragons_rec_j_panel} shows reconstructions with only a subset of levels and Fig.~\ref{fig:dragons_rec_j_panel_44} the contribution made by individual levels.
It can be seen that the error increases gracefully as one disregards finer levels with the reconstructed images becoming smoother but largely without objectionable artifacts, though some ringing can be observed for $j \leq -1$ and $j \leq 0$ in the in-focus region.
Since a reconstruction with a subset of levels is considerably cheaper, requiring only $0.098\%$, $1.72\%$, $13.43\%$ of the computation time for all levels for $j \leq -1$, $j \leq 0$, and $j \leq 1$, respectively, substantial savings for out-of-focus regions are possible when a depth map is known.

\begin{figure*}[t]
  {\includegraphics[width=1.04\columnwidth]{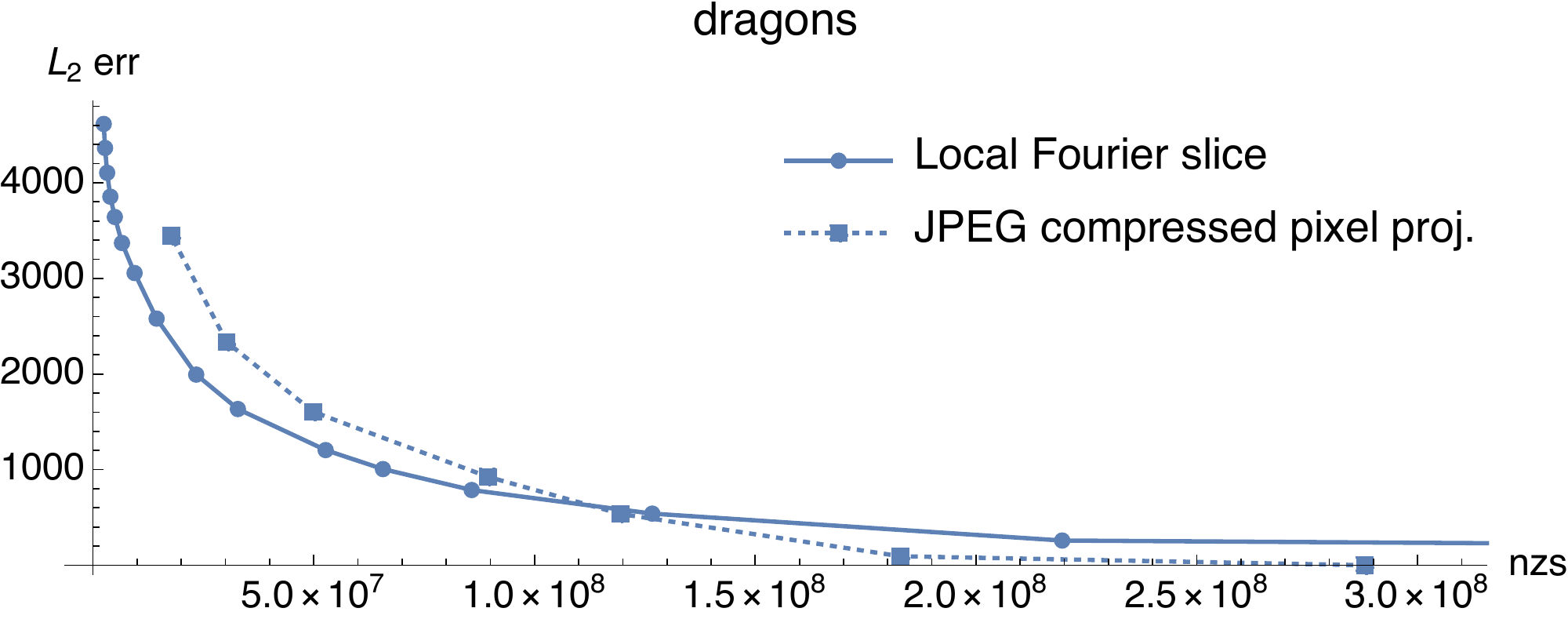}}
  {\includegraphics[width=1.04\columnwidth]{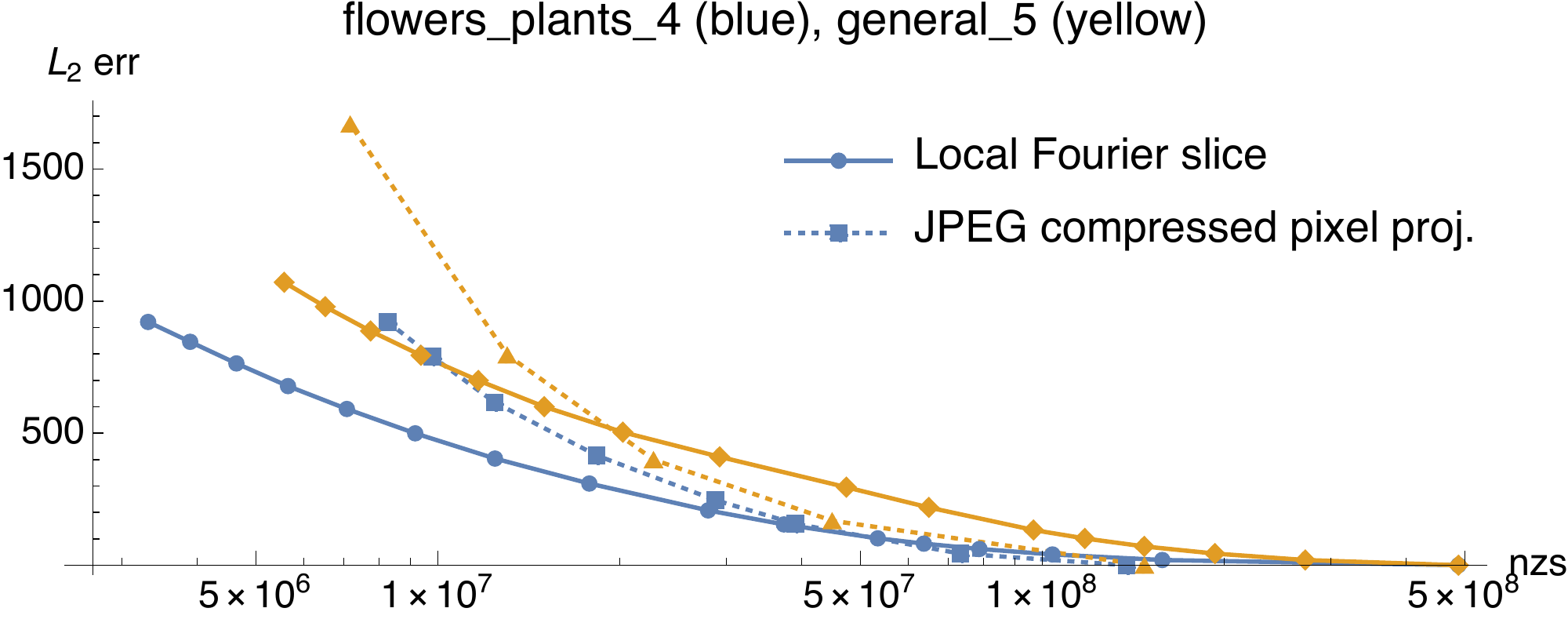}}
  \caption{Comparison of the reconstruction error as a function of the nonzero coefficients in the sparse representation for our technique and the transform coding step of JPEG. Note that for the JPEG-compressed light fields a dense pixel representation is required to reconstruct refocused images. \emph{Left:} Results for synthetic dragon light field with resolution $513 \times 513 \times 33 \times 33 \times 3$ (see Fig.~\ref{fig:dragon_comparison_pixel}). \emph{Right:} Log-linear plot for two representative photographic light fields (flowers\_plants\_4 (fourth column) in Fig.~\ref{fig:panel_general_comp_alpha_flowers_plants} (blue) and general\_5 (fifth column) in Fig.~\ref{fig:panel_general_comp_alpha_general} (yellow)).}
  \label{fig:comp_jpeg}
\end{figure*}

\paragraph{Sparsity}

Fig.~\ref{fig:panel_general_comp_alpha_flowers_plants} and Fig.~\ref{fig:panel_general_comp_alpha_general} show the relationship between the number of nonzero coefficients and $L_1$ reconstruction errors for 10 different photographic light fields for $\alpha = 0.6, 0.85, 1.05, 1.35$ (top to bottom in each figure).
For $\alpha = 0.6$, in both figures one data set (the fifth in Fig.~\ref{fig:panel_general_comp_alpha_flowers_plants} and the third in Fig.~\ref{fig:panel_general_comp_alpha_general}) yields considerably larger errors than the other ones.
For the two data sets large areas with high frequency foliage, requiring many small wavelet coefficients for an accurate representation, are in focus for $\alpha = 0.6$ and this behavior is thus to be expected.
A similar observation holds true for the fourth and fifth data set in Fig.~\ref{fig:panel_general_comp_alpha_general}  although it is less pronounced there since the in-focus regions are smaller.
In contrast, for the third data set in Fig.~\ref{fig:panel_general_comp_alpha_general} one has, except for some edges, a low visual complexity independent of $\alpha$.
Correspondingly, one has a consistently low reconstruction error.
For the remaining data sets there is for each $\alpha$ one region with high visual complexity in focus and the reconstruction errors remain thus relatively constant.
Qualitatively equivalent results hold for the $L_2$ and $L_{\infty}$ norms.
Results for 10 different photographic light fields can be found in the supplementary material.

Fig.~\ref{fig:lf_th} provides a visual comparison of reconstructed images as the sparsity in the light field's wavelet representation increases.
The figure demonstrates that the image fidelity degrades gracefully as the compression rate increases.
Furthermore, when the error becomes visible then it amounts to a lack of high frequency features but largely without objectionable artifacts (image sequences as a function of the compression rate are provided in the supplementary material).
The storage requirements for the coefficients of the compressed light fields in Fig.~\ref{fig:lf_th} (with $cr > 1$) are $2.23$ GB, $1.27$ GB, $660$ MB, $122.39$ MB, and $40.23$ MB, respectively.
As a comparison, the dense light field, required for projection in the pixel domain or Fourier slice photography, requires 13 GB of storage and the uncompressed wavelet representation 48 GB.

In Fig.~\ref{fig:comp_jpeg} we report the $L_2$ error in the reconstructed image as a function of the number of nonzero coefficients for our local Fourier slice photography and a JPEG-compressed representation of the light field (similar to~\cite{Alves2018}).
For the latter, the sparse representation was obtained by considering each slice in the light field (for fixed u-v index) as an image and applying the transform coding step of JPEG, consisting of mask-weighted quantization in the discrete cosine transform (DCT) domain over $8 \times 8$ blocks.
We did not apply JPEG's entropy coding step to obtain a roughly fair comparison to our technique that also only implements transform coding.
Fig.~\ref{fig:comp_jpeg} demonstrates that for large coefficient counts, i.e. little compression, classical JPEG outperforms our approach.
This is to be expected since polar wavelets are redundant and hence more data is required for perfect reconstruction when compared to the non-redundant DCT used in JPEG.
For low coefficient counts, i.e. more aggressive compression, local Fourier slice photography provides, however, lower errors than the JPEG-compressed representations.
Furthermore, and this is at the heart of our work, for the JPEG-compressed light field data sets a dense representation has to be obtained for refocused image reconstruction, requiring GBs of memory, while our approach can perform the projection directly from the sparse wavelet representation.

\paragraph{All-in-focus reconstruction}

Fig.~\ref{fig:allfocus}, right, shows a reconstruction of the dragons scene as well as of the checkerboard ground plane with a depth dependent $\alpha$ value so that the entire scene is in focus.
Slight artifacts are visible around the dragon silhouettes, since we do not take the varying support of the $\zeta_s^{\alpha}(x)$ into account and only sample the depth map at the location of the reconstruction kernels.

\begin{figure}[b]
  \includegraphics[width=\columnwidth]{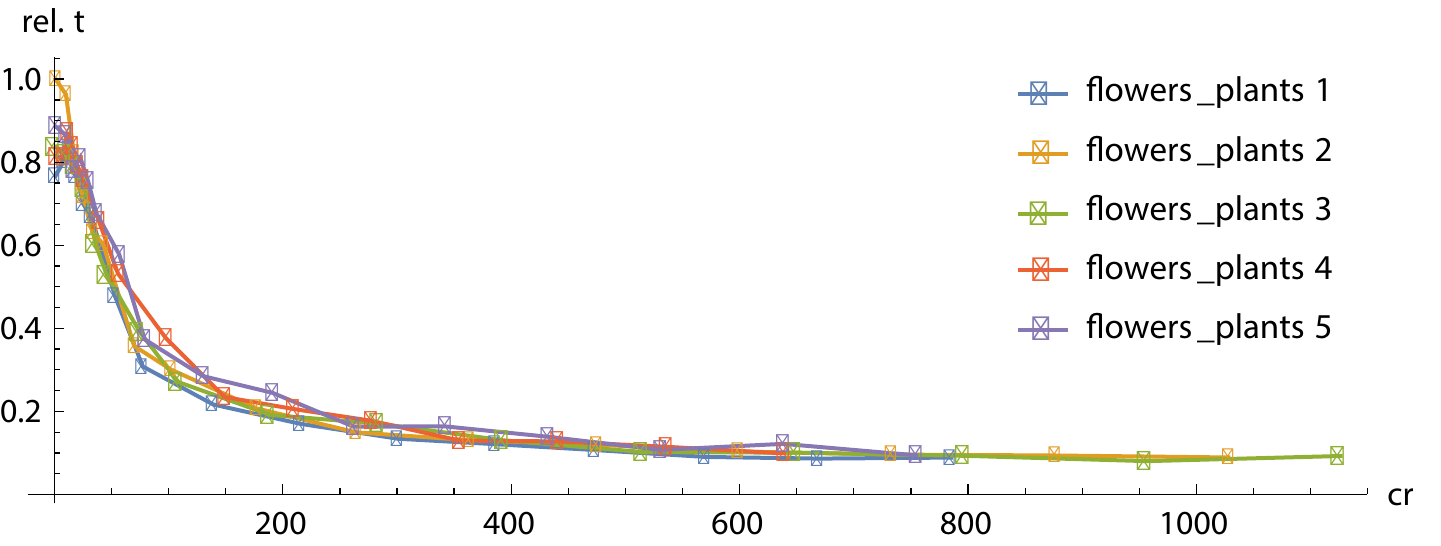}
  \caption{Relative execution time for the light fields in Fig.~\ref{fig:panel_general_comp_alpha_flowers_plants}. The results demonstrate that the use of sparsity can lead to substantially lower computation times. See Fig.~\ref{fig:lf_th} for reconstructed images with the respective compression rates.}
  \label{fig:lf_t_sparsity}
\end{figure}

\begin{figure*}
  \includegraphics[trim=0 60 0 60, clip, width=\textwidth]{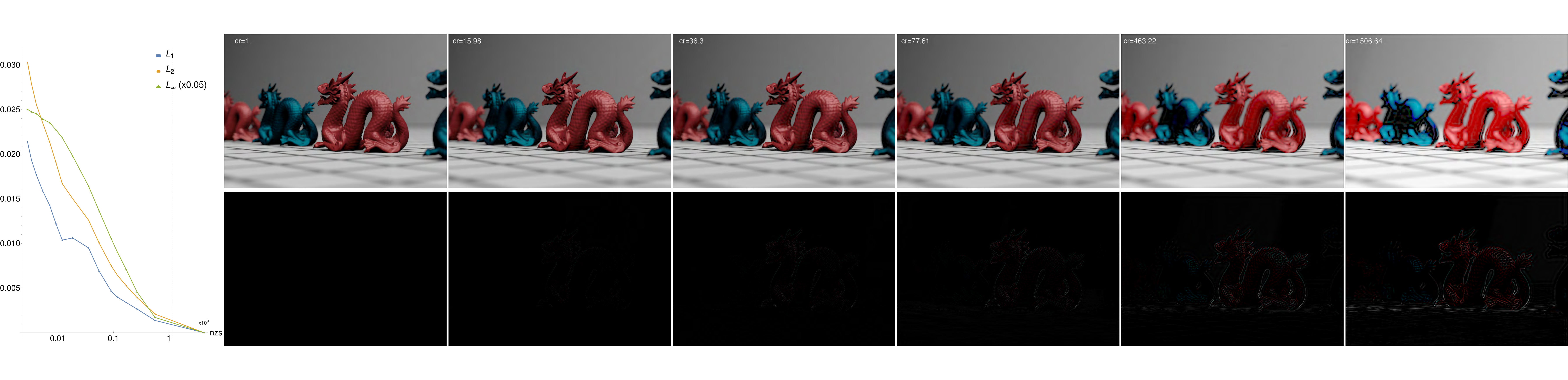}
  \includegraphics[trim=0 60 0 60, clip, width=\textwidth]{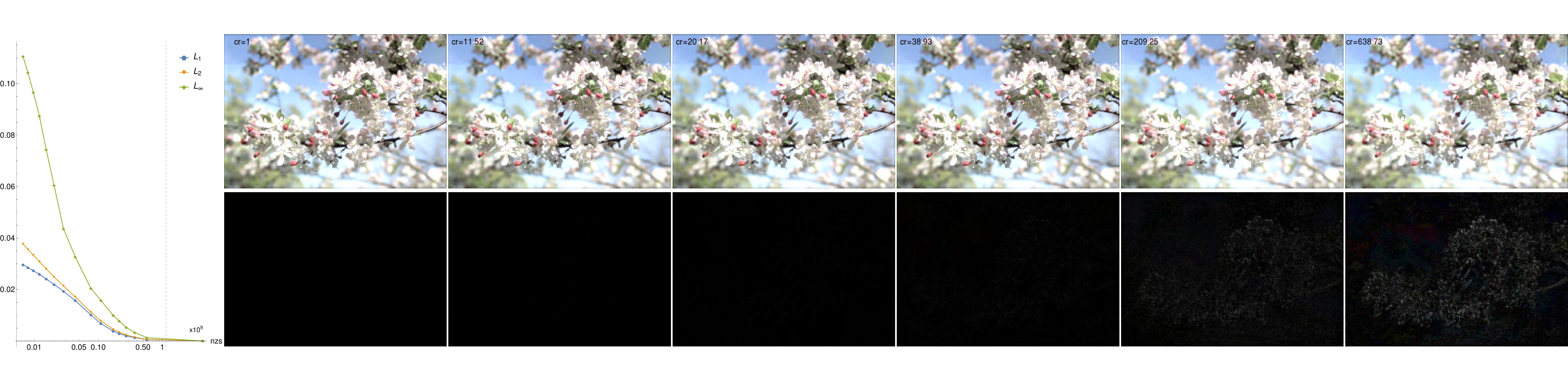}
  \caption{Reconstruction errors as a function of the number of nonzero coefficients in the sparse wavelet representation of the data sets and reconstructed images for $\alpha = 0.9$ for the dragon scene and $\alpha = 1.35$ for the tree blossoms.
  Shown are also difference images compared to the uncompressed reference.
  The results demonstrate that a reconstruction from a sparse wavelet representation of the light fields yields high fidelity images and that even with a very small number of nonzero coefficients, and a correspondingly high compression rates, the errors manifest themselves mainly as missing high frequency details, e.g. on the back of the red dragon, but there are largely no disruptive visual artifacts.
  Animations for the reconstructed image are provided in the supplementary material.}
  \label{fig:lf_th}
\end{figure*}

\paragraph{Performance}

Fig.~\ref{fig:lf_t_sparsity} shows the relative execution time as a function of the compression rate.
Although our implementation is not particularly optimized and we only use the Eigen library for the sparse wavelet representation, the results demonstrate that sparsity can lead to a substantial reduction in computation time.
The plots also show the execution time decreases approximately as $1/\mathrm{cr}$, as one would expect from our analysis of the computational complexity in Sec.~\ref{sec:imaging:complexity}, see in particular Eq.~\ref{eq:complexity:bigo}.

The absolute computation time of image reconstruction is currently approximately two minute for a $1025 \times 1025 \times 33 \times 33 \times 3$ light field data set with a dense wavelet representation and $13$ seconds for the highest compression rate we considered.
The projection in the pixel domain requires in our implementation $25$ seconds from a decompressed, dense light field and roughly the same time is required for the sparse projection with a compression rate of $200$.
The computation of the wavelet representation of the light field requires approximately a minute.

\subsection{Discussion}

Our experimental results demonstrate the practical viability of Algorithm~\ref{algo:reconstruction} for the reconstruction of refocused images from the sparse wavelet representation of a light field.
We verified that high fidelity images can be obtained from a highly sparse representation and that the error increases gracefully with the compression rate.
Furthermore, our experiments show that the error depends on the visual complexity of the in-focus region, which can be exploited when a depth map is available.
Additionally, we demonstrated that simple hard thresholding of polar wavelet coefficients is, at least for moderate to high compression rates, competitive with the transform coding step of JPEG, which uses highly optimized masks.

Our results also show that sparsity in the wavelet representation can lead to a reduction in the computation time through the smaller number of coefficients that has to be processed, although our implementation is currently slower than projection in the pixel domain when a dense representation of a light field is directly available.
Algorithm~\ref{algo:reconstruction} is easily parallelized by exploiting that the reconstruction for each pixel is independent, i.e. there is $N_x \times N_y$ data parallel work.
This makes it directly amenable to a GPU implementation where one could also take advantage of half-precision, which is sufficient to obtain artifact free images.
This would also provide benefits on the embedded processors typically available in cameras.

\begin{figure}[t]
  \includegraphics[width=0.48\columnwidth]{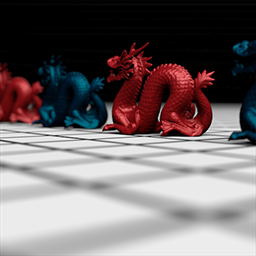}
  \includegraphics[width=0.48\columnwidth]{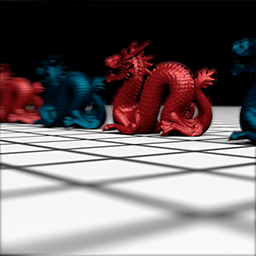}
  \includegraphics[width=0.48\columnwidth]{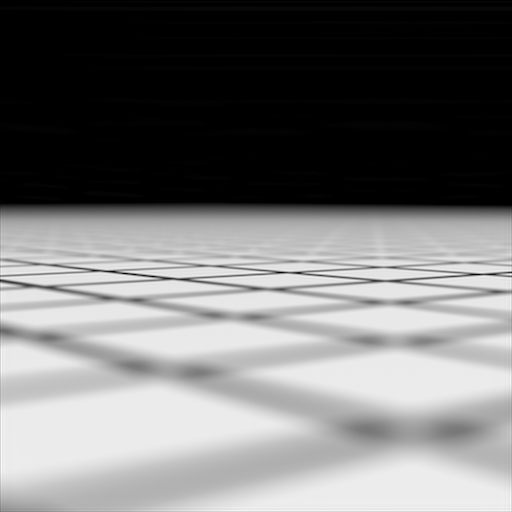}
  \includegraphics[width=0.48\columnwidth]{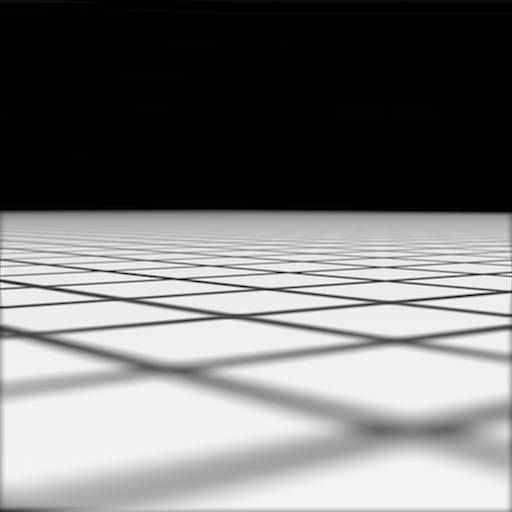}
  \caption{Reconstructions for $\alpha = 1.0$ (left) and with $\alpha$ determined based on a depth map (right) to obtain all-focus reconstructions ($\alpha$ is restricted to $[0.6,1.4]$ which leaves the region nearest to the camera still out of focus). Because we use only a single high resolution depth map and do not consider the variable support of the reconstruction kernels $\zeta_s^{\alpha}$ as a function of level, artifacts are visible around the silhouettes of the dragons.}
  \label{fig:allfocus}
\end{figure}

Fig.~\ref{fig:allfocus} showed first results on the reconstruction of all-focus images from the compressed wavelet representation.
While currently not artifact free when the depth map contains discontinuities, the results verify the potential of our approach  to obtain all-focus images, which is not possible using Fourier slice photography where a fixed $\alpha$ has to be used.
To remove the current artifacts, the depth map needs to be preprocessed in a mip-map-like manner so that an average depth can be sampled at each wavelet level and it might also be necessary to restrict the support of wavelets at depth discontinuities.


\section{Future Work}

The local Fourier slice photography algorithm developed in the foregoing suggests many directions for future work.

In our work, we consider the sparse wavelet representation of light fields, which corresponds to the transform coding step of a compression technique.
To make local Fourier slice photography practical for applications, this has to be extended to a full compression scheme by also considering, for example, blocking, quantization, and color coding.
Our comparison to the transform coding step of JPEG indicates already that the sparse representation will translate to significantly reduced storage requirements also in practice.

Our technique would also benefit from additional work on the wavelet representation.
For example, the radial window $\hat{h}(\vert \xi \vert)$ we currently employ does not provide very good decay in the spatial domain and, similar to~\cite{Ward2015}, one could investigate how better radial windows can be constructed.
One should also further investigate the trade off between increased sparsity for directional, curvelet-like representations and the higher computational costs for evaluating these.
Our results in Sec.~\ref{sec:imaging:slice_alpha_compressed} showed that sparsity in the light field induces sparsity in the reconstructed image.
A quantitative description of this could potentially lead to a further reduction of the storage requirements as well as computation times.

By representing the light field in polar wavelets, many existing techniques for editing and processing are no longer directly available.
However, we believe that many of them can be translated to the polar wavelet domain and it might, in fact, provides advantages to conventional approaches.
For instance, the feature-aware resizing of Gastal and Oliveira~\shortcite{Gastal2017} is naturally formulated in polar coordinates and could hence be performed directly in a sparse representation.
We believe that other tasks, such as the shearlet-based light field reconstruction from a sparse set of views in~\cite{Vagharshakyan2018} or learning based techniques such as~\cite{Kalantari2016} could also benefit from the efficacy of a polar wavelet representation.

In the present work we considered refocused image reconstruction.
Another important application of light fields is novel view synthesis.
It would be interesting to investigate if this can also be performed directly from a sparse polar wavelet representation.

Our current approach for image reconstruction exploits the separability of the refocusing problem so that two dimensional polar wavelets are sufficient.
This is sub-optimal concerning the compression rates that can be attained for the light field and also since one obtains a separable wavelet representation for the reconstructed image.
With 4D polar wavelets, which can be constructed as an extension of the polar wavelets used in the present work, cf.~\cite{Ward2014}, the entire light field could be represented in one wavelet basis and the projection of the data would then yield two dimensional, curvelet-like polar wavelets.
In our opinion, this is both theoretically and practically an interesting direction for future work.

\section{Conclusion}
\label{sec:conclusion}

In this paper, we presented local Fourier slice photography, an algorithm to reconstruct refocused images from a sparse wavelet representation of a light field.
For this, we derived a sheared local Fourier slice equation, which extends the local Fourier slice equation of~\cite{Lessig2018d} to the case of sheared projection.
The equation provides analytic, wavelet-like reconstruction kernels for obtaining a refocused image directly from a light field's wavelet coefficients.
The direct reconstruction from a sparse representation is at the heart of our work and it avoids the need for a dense light field representation that exists for techniques in the literature.
We experimentally verified that high fidelity images can be reconstructed from a highly sparse representations of a light field, providing the potential for significant storage requirements, and our results demonstrate that image quality degrades gracefully as the compression rate increases.
Furthermore, also the computational costs can be reduced by directly reconstructing images from the sparse wavelet representation.
We analyzed this theoretically and demonstrated efficiency gains as a function of compression rate experimentally.
Because of the spatial localization of the wavelets, the costs of our technique depend on the light field's angular resolution. This was not the case for Fourier Slice Photography~\cite{Ng2005} but the localization enables us, for example, to obtain all-focus images, which is not possible using this technique.

%
\begin{acks}
  The comments by the anonymous reviewers helped to considerably improve the manuscript, in particular their insistence to try the algorithm on photographic light fields.
  The Stanford light field group is acknowledged for making the Lytro data sets available to the community.
  First ideas for the project were developed while the author was a post-doc in Marc Alexa's computer graphics group at TU Berlin.
  Many thanks also to Eugene Fiume for continuing support.
\end{acks}

%
\bibliographystyle{ACM-Reference-Format}
\bibliography{fourierslice,pubs,one,lfield,polarlets}

%
\appendix
\section{Spatial representation of reconstruction filters $\zeta_{s}^{\alpha}$}
\label{sec:appendix:zeta:spatial}

When $\hat{h}(\vert \xi \vert)$ is the wavelet window from the steerable pyramid, the $\zeta_{s}^{\alpha,m}(x)$ in Eq.~\ref{eq:slice_alpha:zeta_n:def} are given by
\begin{align*}
  \zeta_{s}^{\alpha,n}(x) &=
  \frac{1}{2 \sqrt{2} x} \pi ^{-\frac{\log (2)+i \pi }{\log (4)}} \left(\left(1-\frac{1}{\alpha }\right)^2 \alpha ^2+1\right)^{-\frac{i \pi }{\log (16)}}
  \\
  & \quad \times \, e^{-i n \left(\pi -\tan ^{-1}\left(\left(1-\frac{1}{\alpha }\right) \alpha \right)\right)} \left(\alpha ^2 x^2\right)^{-\frac{i \pi }{\log (4)}}
  \\
  & \quad \times \, \pi ^{\frac{i \pi }{\log (2)}} A + \left(\left(1-\frac{1}{\alpha }\right)^2 \alpha ^2+1\right)^{\frac{i \pi }{\log (4)}} B
\end{align*}
where
\begin{align*}
  A &= (-i \alpha  x)^{\frac{i \pi }{\log (2)}} (i \alpha  x)^{\frac{i \pi }{\log (4)}} \, \left( e^{i \pi  n} \big(  D_{4}^- - D_{1}^- \big) - \big(  D_{4}^+ - D_{1}^+ \big) \right)
  \\[4pt]
  B &= e^{i \pi  n} (i \alpha  x)^{\frac{i \pi }{\log (4)}} \big(  D_{4}^- - D_{1}^- \big) - (-i \alpha  x)^{\frac{i \pi }{\log (4)}} \big(  D_{4}^+ - D_{1}^+ \big)
\end{align*}
and
\begin{align*}
  D_{d}^{\pm} = \Gamma \left(1 \pm \frac{i \pi }{\log (4)}, \pm \frac{i \pi  x \alpha }{d \sqrt{\left(1-\frac{1}{\alpha }\right)^2 \alpha ^2+1}}\right) .
\end{align*}

\end{document}